\documentclass[extra]{gji}
\usepackage[]{graphicx}
\usepackage{subeqn,url,times}
\newcommand{\srule}{\rule[0mm]{0mm}{1.25em}}

\newcommand{\sst}{\scriptstyle}

\newcommand{\Trm}{^{\rm{\sst{T}}}}

\newcommand{\atan}{{\mathrm{atan}}}
\newcommand{\mF}{{\mathcal{F}}}
\newcommand{\mJ}{{\mathcal{J}}}
\newcommand{\mS}{{{S}}}
\newcommand{\mW}{{{A}}}
\newcommand{\mT}{{\mathcal{T}}}
\newcommand{\mU}{{\mathcal{U}}}
\newcommand{\mP}{{\mathcal{P}}}
\newcommand{\bm}{{\mathbf{m}}}
\newcommand{\bmh}{{\mathbf{\hat{m}}}}
\newcommand{\bdh}{{\mathbf{\hat{d}}}}
\newcommand{\bx}{{\mathbf{x}}}

\newcommand{\bn}{{\mathbf{n}}}
\newcommand{\bd}{{\mathbf{d}}}
\newcommand{\bw}{{\mathbf{w}}}
\newcommand{\bwh}{{\mathbf{\hat{w}}}}
\newcommand{\bnabla}{\mbox{\boldmath$\nabla$}}

\newcommand{\bK}{{\mathbf{K}}}
\newcommand{\bS}{{\mathbf{S}}}

\newcommand{\hsps}{\hspace*{0.75em}}
\newcommand{\be}{\begin{equation}}
\newcommand{\ee}{\end{equation}}
\newcommand{\ber}{\begin{eqnarray}}
\newcommand{\eer}{\end{eqnarray}}

\date{Accepted \today. Received \today; in original form \today}
\pagerange{\pageref{firstpage}--\pageref{lastpage}}
\volume{XXX}
\pubyear{2011}

\begin{document}
\title[Wavelets, Sparsity, and Seismic Tomography]
{Solving or resolving global tomographic models with spherical
  wavelets, and the scale and sparsity of seismic heterogeneity}
\author[Simons, Loris, Nolet, Daubechies \& others]{
Frederik J.~Simons$^1$, Ignace Loris$^2$, Guust Nolet$^3$, 
Ingrid C.~Daubechies$^4$,   
\vspace{0.25em}\\
\LARGE{\rm{S.~Voronin$^4$, J.~S.~Judd$^4$, P.~A.~Vetter$^4$,
J.~Charl\'ety$^3$ and C.~Vonesch$^4$}}\\  
$^1$ Department of Geosciences, Princeton University, Princeton,
NJ 08544, USA. E-mail: fjsimons@alum.mit.edu\\
$^2$ Mathematics Department, Universit\'e Libre de Bruxelles, CP 217, 
Boulevard du Triomphe, 1050 Brussels, Belgium.\\ 
$^3$ G\'eoAzur, Universit\'e de Nice, 06560 Sophia Antipolis,
France\\
$^4$ Program in Applied and Computational Mathematics, Princeton
University, Princeton, NJ 08544, USA
} 
\maketitle

\begin{summary}
  We propose a class of spherical wavelet bases for the analysis of
  geophysical models and for the tomographic inversion of global
  seismic data. Its multiresolution character allows for modeling with
  an effective spatial resolution that varies with position within the
  Earth. Our procedure is numerically efficient and can be implemented
  with parallel computing.  We discuss two possible types of discrete
  wavelet transforms in the angular dimension of the cubed sphere.  We
  discuss benefits and drawbacks of these constructions and apply them
  to analyze the information present in two published seismic
  wavespeed models of the mantle, for the statistics and power of
  wavelet coefficients across scales. The localization and sparsity
  properties of wavelet bases allow finding a sparse solution to
  inverse problems by iterative minimization of a combination of the
  $\ell_2$~norm of data fit and the $\ell_1$~norm on the wavelet
  coefficients. By validation with realistic synthetic experiments we
  illustrate the likely gains of our new approach in future inversions
  of finite-frequency seismic data and show its readiness for global
  seismic tomography.
\end{summary}
\begin{keywords}
Inverse problems, seismic tomography, sparsity, spectral analysis, wavelets
\end{keywords}

\label{firstpage}


\section{I~N~T~R~O~D~U~C~T~I~O~N}

As long as tomographic Earth models remain the solutions to
mixed-determined inverse problems \cite[]{Nolet87,Nolet2008} 
there will be disagreement over the precise location, shape, form and
amplitude of lateral and radial anomalies in seismic wavespeed that
exist within the Earth; there will be attempts to derive the
best-fitting mean structure \cite[e.g.][]{Becker+2002}, and the needed
efforts to validate them \cite[e.g.][]{Qin+2009,Bozdag+2010}. At the
same time, patterns, second-order structure and correlations  between
and within models will continue to be sought with the goal of
characterizing seismic heterogeneities
\cite[e.g.][]{Passier+95,Bergeron+99,Hernlund+2008} or
relating them to geochemical \cite[e.g.][]{Gurnis86}, tectonic
\cite[e.g.][]{Yuen+2002,Becker2006}, or geodynamical
\cite[e.g.][]{Jordan+93,Piromallo+2001,Houser+2009}
processes. It has also become clear that model
characteristics such as the power spectrum of tomographic anomalies
\cite[][]{Chevrot+98a,Chevrot+98b,Boschi+99}
may teach us as much about the modeler's choices of parameterization
and regularization as about the model, without imparting much
information about the physical or statistical nature of our complex,
physically and chemically differentiated system Earth --- yet the
latter should be our target. In recent work questions about the size
and scale distribution of Earth structure have  more
fruitfully been addressed by direct inference from the data themselves
\cite[e.g.][]{Hedlin+2000,Margerin+2003,Becker+2007,Garcia+2009}
without the detour of first deriving a global three-dimensional model
and analyzing that.

By no means are the analysis and representation of volumetric
properties the sole purview of seismology or geodynamics, and thus it
is not surprising that there is a large literature on the subject in
virtually every area of scientific inquiry (e.g. medical imaging,
astronomy, cosmology, computer graphics, image processing,~...). In
this context much has come to be expected of the special powers of
wavelets, with their built-in discriminating sensitivity to structure in
the space and spatial frequency domains
\cite[]{Daubechies92,Strang+97,Mallat2008}. Notwithstanding a
continued interest and clear and present progress in the field
\cite[e.g.][]{Foufoula+94,Klees+2000,Freeden+2004b,Oliver2009},
the use of wavelets is still no matter of routine in the geosciences,
beyond applications in one and two Cartesian dimensions. This
despite, or perhaps because, there being a wealth of available
constructions relevant for global geophysics, in other words: on the
sphere \cite[e.g.][]{Schroder+95,Narcowich+96,Antoine+2002,
Holschneider+2003,Freeden+2004a,Hemmat+2005,Schmidt+2006,Starck+2006,
McEwen+2007,Wiaux+2007,Lessig+2008}, if not on the ball. Indeed,
inasmuch as they involve the analysis of cosmological data or
computer-generated images, the above studies are mostly concerned with
surfaces, not volumes.

In seismology, \cite{Chiao+2001} were, to our knowledge, the first to
develop a ``biorthogonal-Haar'' wavelet lifting scheme
\cite[]{Schroder+95} for a triangular surface tesselation of the
sphere suitable for multiscale global tomography. Later, these same
authors formed a (biorthogonal) spline basis for a Cartesian cube
useful in exploration geophysics \cite[]{Chiao+2003} and for regional
studies \cite[]{Hung+2010}. Finally, \cite{Chevrot+2007} constructed a
three-dimensional (orthogonal) Haar basis on an equidistant
geographical grid which was also used for a regional inversion. To
this date, a truly three-dimensional  wavelet basis on the
ball with practical utility in the geosciences has been lacking.

Whatever the role that wavelets will play in it, the future of global
seismic tomography will continue to involve massive amounts of
heterogeneous data spanning a range of resolutions, from the travel
times reported by the global networks to the waveforms of portable
deployments, with strong regional concentrations of station coverage
in areas such as Japan, the United States, and Europe, supplemented
with sparse or non-existent networks in less densely populated or
oceanic regions. It is also clear that finite-frequency kernels, which
allow for the correct volumetric sensitivity-based weighting of the
measurements in distinct frequency bands, are here to stay, whichever
the various ways in which they are calculated \cite[see][and
references therein]{Nolet2008}. Accounting for finite-frequency
sensitivity requires an effective overparameterization if one wishes
to exploit the extra resolution offered by the spatial variations in
their sensitivity, something for which wavelet seem ideally suited
also. 

This paper documents some of the extensive prospective work that we
have done in preparation for realistic wavelet-based global seismic
inversions. We wish to share the most important insights that we have
gained through these studies. Our goal remains to ensure that there
exist performant and efficiently calculable, flexible wavelet methods
on the three-dimensional ball, to fulfill the promise of
multiresolution analysis \cite[][]{Mallat89a,Jawerth+94} in global
seismology. Not just for the representation and analysis of seismic
models after the fact, but rather for their determination, as an
integral part of a parsimonious parameterization of the inverse
problem --- of the sensitivity matrix, of the model space, or both.
Although there is no objective guarantee that Nature, or the
interior of the Earth in particular, is parsimonious in character,
sparsity is worth striving for. By simplifying a tomographic image to
contain a relatively small number of recognizable objects we
facilitate interpretation~\cite[]{Sambridge+2006}. Moreover, such
models can be more accurate than their data~\cite[]{Gauch2003}, a
point not to be overlooked in view of the large relative errors of
seismic delay times and amplitudes.

By \textit{flexibility} we mean the ability to substitute a particular
wavelet design for another in any of the three coordinate directions;
by \textit{efficiency} we intend to avoid the tedious case-by-case
derivation of different bases and calculation methods. By
\textit{performance} we target the ability to capture the unknown
model by explaining the data (in an $\ell_2$~sense) with a minimum of
wavelet and scaling coefficients, both where the data require the
solution to be smooth and where they necessitate the presence of sharp
contrasts. It is of course in this capacity also
\cite[e.g.][]{Donoho+94,Donoho+95b} that wavelets will distinguish
themselves from any other traditional method of seismic inversion. As
to \textit{sparsity}, it is both numerically and philosophically
attractive \cite[]{Constable+87} and physically plausible or at least
testable that the interior of the Earth should be sparse when
expressed in a wavelet basis. Fortunately, for most large
underdetermined systems of linear equations the minimal $\ell_1$~norm
solution is also the sparsest \cite[]{Candes+2006,Donoho2006}, for
which (fast, iterative) algorithms exist
\cite[]{Daubechies+2004,Loris2009}. Elsewhere,
\cite{Loris+2007,Loris+2010} and \cite{Gholami+2010} explored the
suitability of sparsity-seeking thresholded wavelet-based inversion
approaches in two-dimensional (2D) and three-dimensional (3D)
Cartesian settings relevant to seismic tomography. All of the above
issues will again be the guiding principles behind the new spherical
wavelet construction(s) that we present in this paper.

\begin{figure}\centering
\includegraphics[width=0.85\columnwidth]{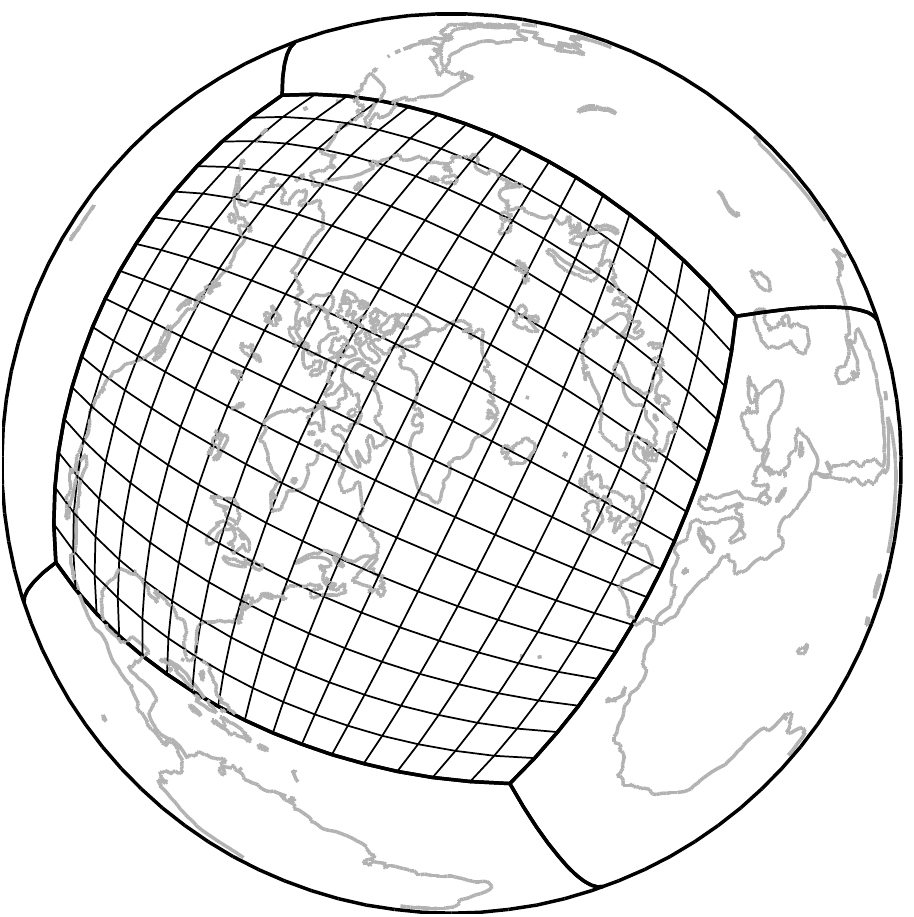}
\caption{\label{thechunk}Aerial view showing our first adaptation of
the cubed sphere of \cite{Ronchi+96}. Of the front-facing four of
the in total six ``chunks'', one is gridded to reveal its
$2^{2N}$ distinct surface elements ($N=4$).}
\end{figure}
This paper is organized as follows. In
Section~\ref{firstconstructionsect} we develop a first class of
wavelet constructions on the sphere via a well-known
Cartesian-to-spherical mapping known as the ``cubed sphere''
\cite[]{Ronchi+96,Komatitsch+2002a}. As this surface tesselation has
``seams'' separating each of six subdivisions or ``chunks'', we
acknowledge these boundaries in the construction by using so-called
``wavelets on the interval''. These revert to the classical compactly
supported (bi)orthogonal Cartesian constructions of
\cite{Daubechies88b} and \cite{Cohen+92} in the interior domains but
receive special consideration on the edges as put forth
by~\cite{Cohen+93}.  In Section~\ref{tomosparse} we study the sparsity
of two global seismic tomographic Earth models by thresholded
reconstructions of their wavelet transforms applied to the angular
coordinates of the cubed sphere, at constant depth intervals, and
considering a variety of goodness-of-fit criteria.
\begin{figure*}
\includegraphics[width=0.75\textwidth]{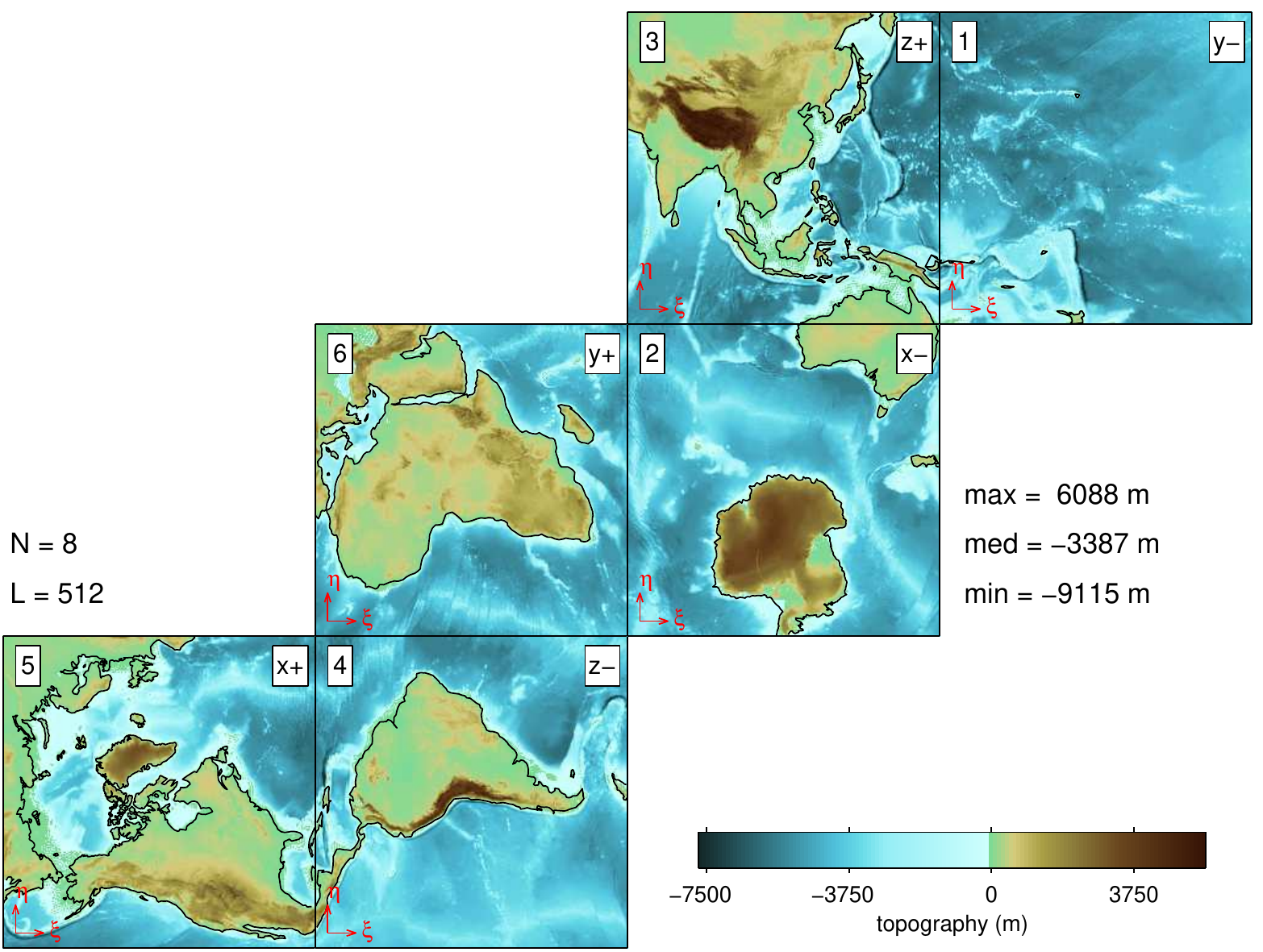}
\caption{\label{twodtopo}Geometry, nomenclature, and numbering of the
six faces of our first adaptation of the cubed sphere of
\cite{Ronchi+96} in a two-dimensional ``unfolded'' view. Rendered is
the Earth's topography from the model ETOPO5, courtesy of NOAA's
National Geophysical Data Center. The projection was obtained by
spherical-harmonic expansion of the coefficients from this model
(Georg Wenzel, \textit{pers.~comm.}) truncated at degree and order
$L=2^{N+1}$, evaluated at the $6\times 2^{2N}$ cubed-sphere grid points
$\xi,\eta$, for $N=8$. Minimum, median, and maximum values in this
approximation are shown in the legend.}\vspace{-1em}
\end{figure*}
We furthermore characterize, in Section~\ref{tomosparse2}, the scale
lengths of heterogeneity in these models by reporting the relative
contributions of their wavelet and scaling coefficients in the
expansion. In Section~\ref{inverseproblem} we review the main approach
to obtain sparse wavelet-based solutions to the \textit{inverse}
problem of seismic tomography, which were previously discussed in a
Cartesian framework by \cite{Loris+2007,Loris+2010}. As using the
initial construction with such schemes led to undesirable artifacts at
the edges between the chunks, we derive a second wavelet construction
in Section~\ref{secondconstructionsect}, which appears to be free of
such artifacts, as we show using realistic synthetic tests in
Section~\ref{experiment}. While in this paper we focus on the angular
part of the cubed \textit{sphere} we generalize our construction to
including the case of the \textit{ball} and provide an outlook for
further research in global seismic tomography in the concluding
Section~\ref{conclusions}.

\section{A{\hsps}F~I~R~S~T{\hsps}C~O~N~S~T~R~U~C~T~I~O~N}

\label{firstconstructionsect}

As \cite{Ronchi+96}, we define the coordinate
quartet~$(\xi,\eta,r,\kappa)$ for each of the $\kappa=1\rightarrow 6$
chunks. The angular coordinates $-\pi/4\leq \xi,\eta \leq\pi/4$ and
the radial coordinate~$r$ are mapped to the usual Cartesian
triplet~$(x,y,z)$ using the transformation
\be\label{xieta2xyz}
(x,y,z)=\left\{\!\!
\begin{array}{lcl}
r\, (\tan\eta , -1  , -\tan\xi )/s & \mathrm{if} & \kappa=1, \\
r\,(-1 , -\tan\xi , \tan\eta )/s & \mathrm{if} & \kappa=2, \\
r\,(\tan\eta , -\tan\xi , 1  )/s & \mathrm{if} & \kappa=3, \\
r\,(-\tan\xi , \tan\eta  , -1 )/s & \mathrm{if} & \kappa=4, \\
r\,(1 , \tan\eta , -\tan\xi )/s & \mathrm{if} & \kappa=5, \\
r\,(-\tan\xi, 1 , \tan\eta )/s & \mathrm{if} & \kappa=6,\\
\end{array}
\right.
\ee
whereby~$s=\sqrt{1+\tan^2\xi+\tan^2\eta}$. 
The inverse mapping is obtained, for~$t=\max(|x|,|y|,|z|)$, 
\be\label{xyz2xieta}
(\xi,\eta,\kappa)=\left\{\!\!
\begin{array}{lcl}
{}[\,\atan(z/y), \atan(-x/y), 1\,] & \mathrm{if} & t=-y,\\
{}[\,\atan(y/x), \atan(-z/x), 2\,] & \mathrm{if} & t=-x,\\
{}[\,\atan(-y/z), \atan(x/z), 3\,] & \mathrm{if} & t=z,\\
{}[\,\atan(x/z), \atan(-y/z), 4\,] & \mathrm{if} & t=-z,\\
{}[\,\atan(-z/x), \atan(y/x), 5\,] & \mathrm{if} & t=x,\\
{}[\,\atan(-x/y), \atan(z/y), 6\,] & \mathrm{if} & t=y,\\
\end{array}
\right.
\ee
whereby $r=\sqrt{x^2+y^2+z^2}$. This parametrization is non-smooth
across the edges separating the chunks. The above formulas correspond
to the drawing in Fig.~\ref{thechunk}, where only one of the chunk
faces is gridded to reveal the angular coordinate lines $(\xi,\eta)$
at a resolution that divides this face into $2^4\times 2^4$ distinct
surface elements. Throughout this paper we will quote $N$ as the
angular resolution level of our cubed sphere, which implies that it
has~$6\times 2^{2N}$  such elements, with typical tomography grids
having $N=7$.

In principle there are many possibilities to choose the surficial
coordinates  $(\xi,\eta)$ in each chunk. We picked ours so as to
minimize the splitting of continents over more than one chunk. Our
choice differs from the canonical version of \cite{Ronchi+96} by a
rigid rotation of the coordinate system, as can be seen by comparing
our Fig.~\ref{twodtopo} with their Figs~15--16. The Euler angles
used in our construction are $\alpha=0.0339$,
$\beta=1.1705$, and $\gamma=1.1909$, respectively. It is important to
note that within a chunk $\xi$ and $\eta$ are \emph{not} spherical 
coordinates; a shift in $\xi$ (with $\eta$ fixed) or in
$\eta$ (with $\xi$ fixed) does \emph{not} correspond to a
rotation on the sphere. This is apparent from the pinching of
coordinate lines in Fig.~\ref{thechunk}.

\begin{figure*}
\includegraphics[width=\textwidth]{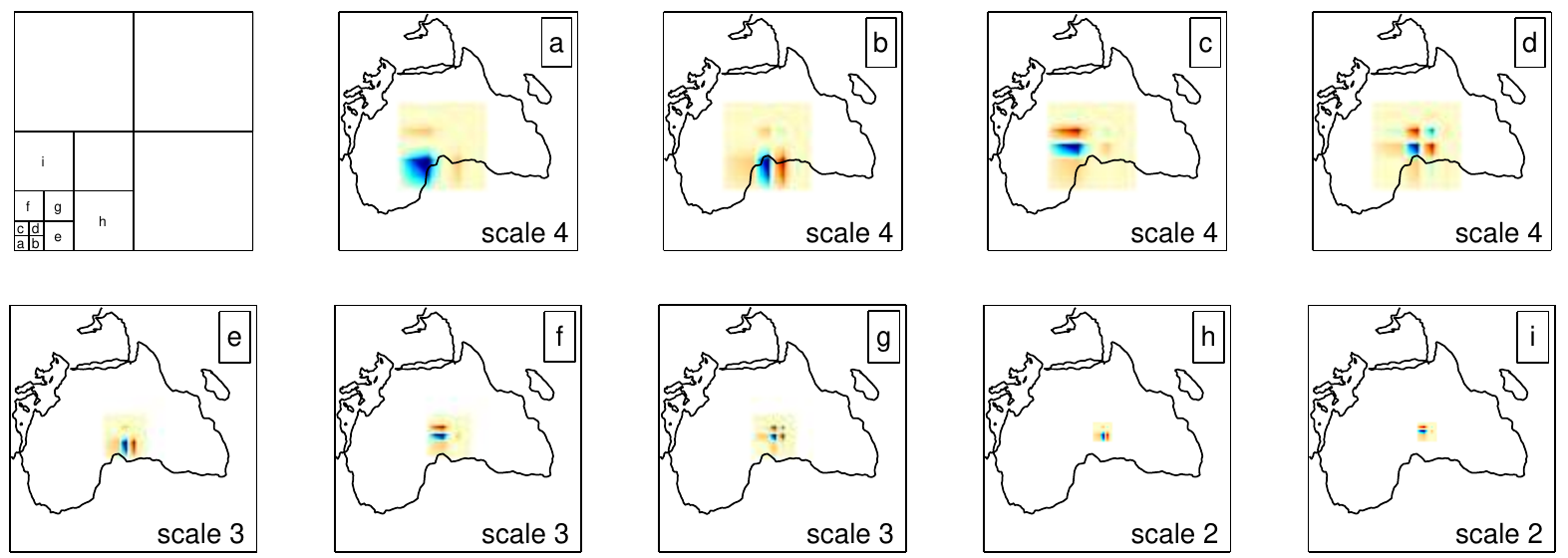} 
\caption{\label{scalesD4}Wavelet and scaling functions of the D4
  construction in the angular coordinates of the cubed sphere, at
  various scales. The positions of the coefficients belonging to the
  functions in the lettered panels are shown in the diagram in the top
  left. The scaling function~(\textit{a}), which is averaging in
  nature, captures what remains to be explained after the breakdown
  into wavelets down to scale 4 is complete. Each of the wavelets,
  which pick up detailed, derivative, structure, is sensitive in a
  particular direction: to $\xi$ in~(\textit{b}), to $\eta$
  in~(\textit{c}), or diagonally in~(\textit{d}).  In the interior
  domain, away from the edges where boundary functions (not shown)
  live, the patterns repeat exactly, with the footprint at each
  successive scale half that of the preceding scale. The diagonally
  sensitive wavelet at scale~2 is not shown. Every function shown is
  orthonormal in ($\xi,\eta$) and their inner products with respect to
  every other one vanish.}
\end{figure*}
\begin{figure*}
\includegraphics[width=\textwidth]{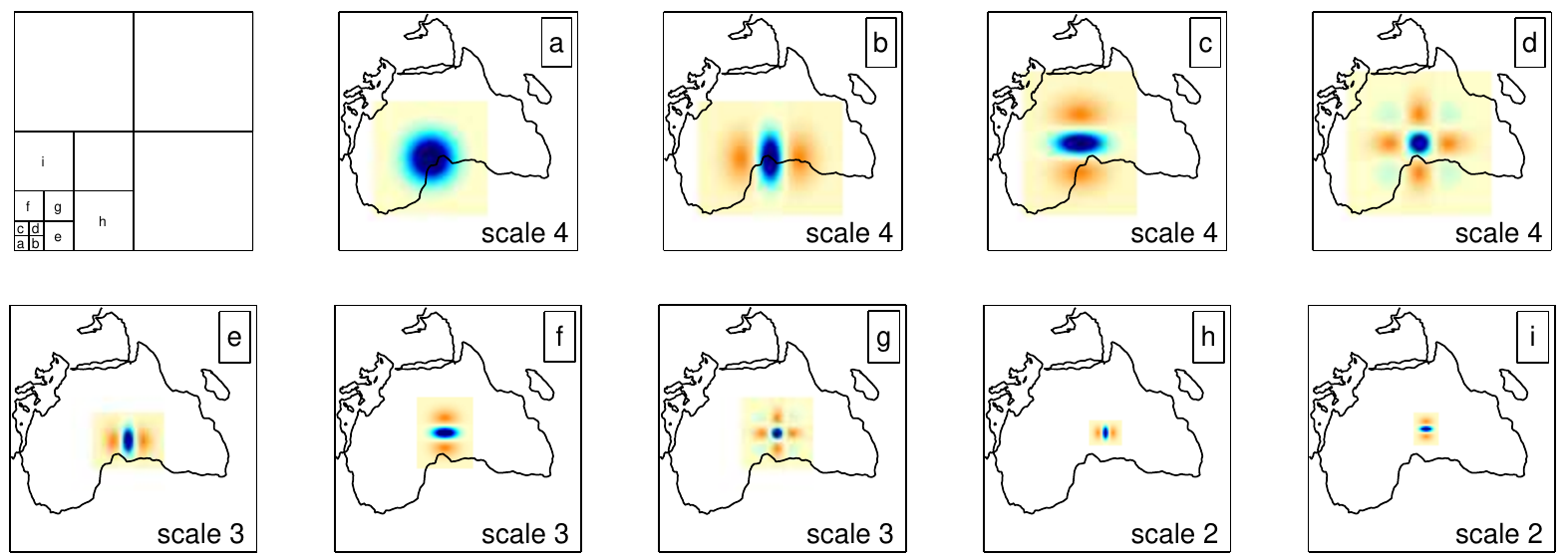}
\caption{\label{scalesCDF}Wavelet and scaling functions arising from the
CDF 4--2 construction in the angular coordinates of the cubed sphere,
with their scale levels indicated. The layout is identical to that of
Fig.~\ref{scalesD4}. As opposed to the D4 wavelets, the CDF 4--2
construction is biorthogonal, which renders every shown synthesis
function orthogonal in ($\xi,\eta$) to its dual, which is used for
analysis; none of the dual functions are shown. Unlike the D4
functions the CDF 4--2 have mirror symmetry. 
}
\end{figure*}

Armed with the coordinate conversions of eqs~(\ref{xieta2xyz})
and~(\ref{xyz2xieta}) we are able to regard the problem of designing a
wavelet transform for the sphere as simply requiring the selection of
a certain Cartesian wavelet transform which is mapped to and from the
sphere. Such an approach is philosophically related to those involving
stereographic projection~\cite[]{Antoine+99,Antoine+2002,Wiaux+2005},
though the fundamental domain of our transform remains a single chunk.
Within each such chunk, the surface Jacobian of our mapping is given
by the smoothly varying
\be
\mJ(\xi,\eta)=(1+\tan^2\!\xi)(1+\tan^2\!\eta)/s^3
,\quad
\sqrt{2}/2\le J\le1.
\ee
For each of the chunks then the area is given by
\be
\int_{-\pi/4}^{\pi/4}\int_{-\pi/4}^{\pi/4}\mJ(\xi,\eta)\,d\xi\,d\eta=
\frac{4\pi}{6}
.
\ee
Without this being a uniform mapping, one of the main advantages of the
chosen coordinate system is thus that the meshes defined on each
region span the surface of the sphere with an almost constant spatial
resolution, as noted by~\cite{Ronchi+96}.

Ignoring any and all such distortions we are able to unlock the power
of popular Cartesian wavelet constructions, of which we choose the two
best known: the orthogonal construction of \cite{Daubechies88b}
and the biorthogonal construction of \cite{Cohen+92}. Both of these
lead to compactly supported wavelets and scaling functions, though
only the biorthogonal ones can be (anti)symmetric (except for Haar).
Examples of scaling functions and wavelets at scales of decreasing
dominant wavelength are shown in Fig.~\ref{scalesD4} for the four-tap
Daubechies basis (D4) and in Fig.~\ref{scalesCDF} for the
Cohen-Daubechies-Feauveau family with four and two vanishing moments
(CDF 4--2) in analysis and synthesis, respectively.

\begin{figure*}
\includegraphics[width=0.75\textwidth]{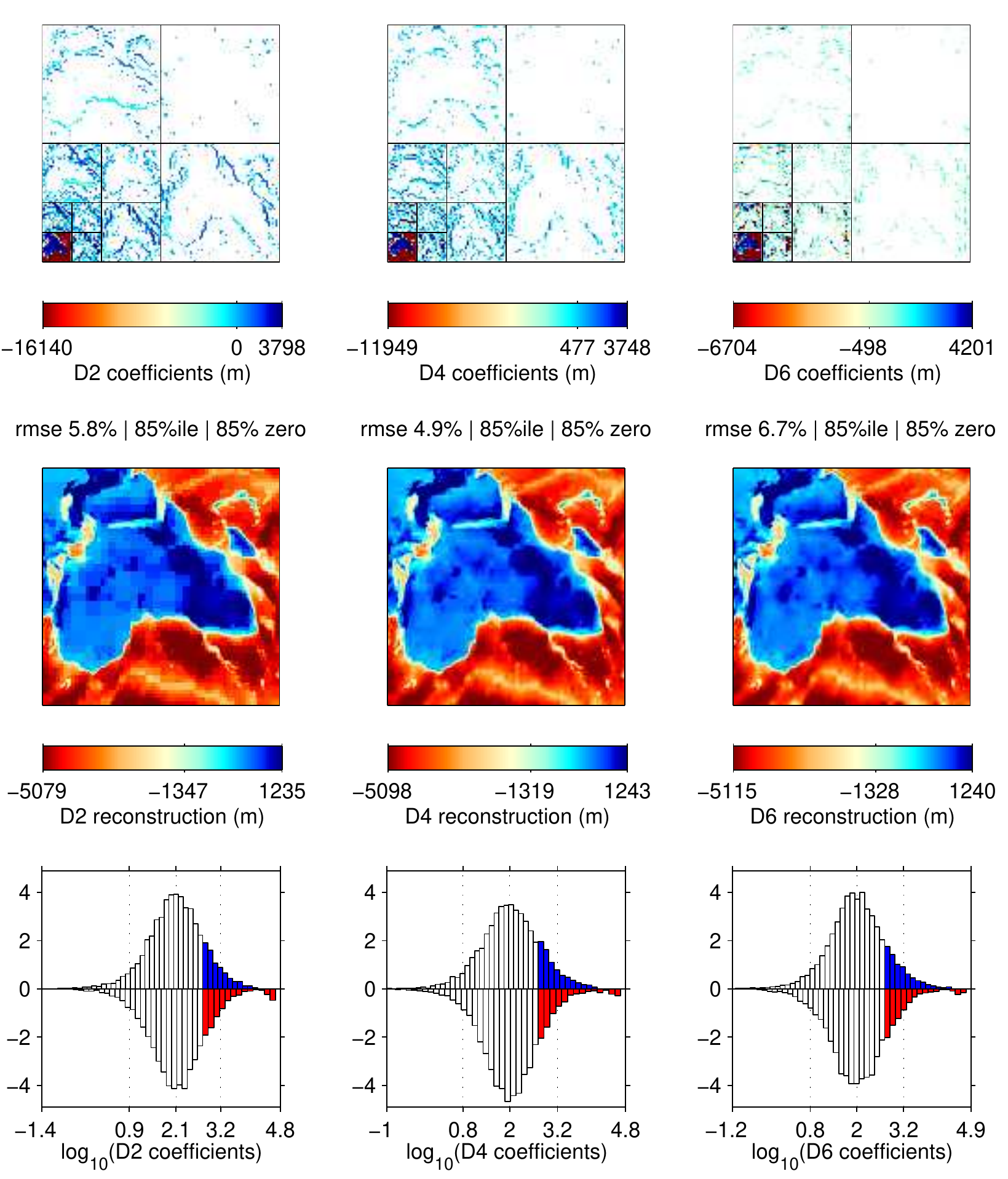} 
\caption{\label{wavelettopo} Wavelet and scaling coefficients
  (\textit{top}), space-domain reconstructions after thresholding
  (\textit{middle}), and ``signed'' histograms (\textit{bottom}) of
  the wavelet and scaling coefficients of the ``African'' (sixth) face
  of the cubed-sphere version of the Earth's topography first shown
  in Fig.~\ref{twodtopo}. We have used the preconditioned
  interval wavelet transforms on the faces of the cubed sphere, as
  described in the text. All coefficients were hard-thresholded at the
  $85^\mathrm{th}$ percentile level, retaining only the 15\% largest
  coefficients by absolute value. In the top row, the locations of
  zeroed coefficients are rendered white; those are also captured by
  the white bars in the histograms.  The root mean squared (rms) error
  of the reconstruction after thresholding is indicated as a
  percentage of the signal rms. Tick marks on the color bars identify the
  $5^\mathrm{th}$, $50^\mathrm{th}$ and $95^\mathrm{th}$ percentile of
  the coefficients or the spatial reconstructions after thresholding,
  respectively. Interior ticks on the histograms roughly coincide with
  these same percentiles as applied to either the positive and
  negative coefficients when expressed on a logarithmic scale.
  Histograms for the positive coefficients point up and have ordinates
  in positive percentages, histograms for the negative coefficients
  point down and have ordinates in negative percentages; these
  percentages are with respect to the total number of positive and
  negative coefficients. The blue and red shaded areas of the
  histograms reflect the coefficients retained at the global
  $85^\mathrm{th}$ thresholding level.}
\end{figure*}

The literature on Cartesian wavelet analysis is vast, and it is not
our intention to repeat any of it here. Most useful for the practicing
geophysicist will perhaps be the treatises by \cite{Mallat2008} and
\cite{Strang+97}; texts focused on algorithms are \cite{Press+92} and,
in particular, \cite{Jensen+2001}. All of the computer code required
to reproduce the figures and conduct the analyses presented in this
paper is molded after these general references and will be
available from the authors.  

Two aspects of wavelet analysis bear specific mentioning here. The
first intricacy is how we treat the seams between the chunks. In
agreement with \cite{Cohen+93} the argument is easily made that
neither ignoring the seams nor periodization or reflection are
viable options, as each of these leads to artifacts in the
representation. We thus follow their suggestion to the letter and
construct a multiresolution basis requiring $2^{2N}$ wavelet and
scaling coefficients for each of the chunk faces having $2^{2N}$
surface elements. For this we switch to special boundary filters at
each of the edges, and apply preconditioners to the data prior to
transformation in order to guarantee the usual polynomial cancellation
throughout the closed rectangular interval. The acknowledgment of the
edges in this way is the hallmark of the wavelet construction in this
section here (which we call the First Construction). This is as easily
done for the orthonormal as for the biorthogonal constructions, though
we have limited the implementation and illustration of this procedure,
in  Fig.~\ref{wavelettopo}, to the compactly supported two-tap (Haar),
four-tap, and six-tap orthonormal families (D2, D4 and D6).  

Before we discuss Fig.~\ref{wavelettopo} in any more detail we should
introduce the second important feature that renders wavelet transforms
in general useful for the analysis and representation of (geophysical)
data. This second topic is the idea of \textit{thresholding}, or
shrinkage. In many applications the wavelet transformation amounts to
a projection under which many of the expansion coefficients are very
small: so small that we might as well throw them away; the resulting
reconstruction will still be close to the original~\cite[]{Donoho+94}.
Intuitively, the ``best'' wavelet basis that we can select to
represent our data is the one that yields the most near-zero
coefficients. When these are \textit{replaced} by zeroes prior to
reconstruction, as under the definition of \textit{hard}
thresholding~\cite[]{Mallat2008}, we obtain highly compressed versions
of the data at hand, with only negligible degradation.

Fig.~\ref{wavelettopo} explores the effects of thresholding,
coefficient statistics, and reconstruction errors for a model of
terrestrial topography, a general proxy for the length scales of
heterogeneities to be found not only at the surface, but also in the
interior of the Earth. We focus on the sixth, or ``African'' chunk of
our cubed sphere, and use the D2, D4 and D6 wavelet bases (on the
interval, with preconditioning). The top row uses the (common)
conventions introduced in Fig.~\ref{scalesD4} in plotting the wavelet
and scaling coefficients in each of the basis after (hard)
thresholding them such that only the coefficients larger than their
value at the 85$^\mathrm{th}$ percentile level survive. The
coefficients that have now effectively been zeroed out are left white
in these top three panels. The middle series of panels of
Fig.~\ref{wavelettopo} plots the spatial reconstruction after
thresholding at this level; the root mean squared (rms) error of
these reconstructions are quoted as a percentage of the original root
mean squared signal strength. The thresholded wavelet transforms allow
us to discard, as in these examples, 85\% of the numbers required to
make a map of African topography in the cubed-sphere pixel basis: the
percentage error committed is only 5.8\%, 4.9\% and 6.7\% according to
this energy criterion in the D2, D4 and D6 bases, respectively. From
the map views it is clear that despite the relatively small error, the
D2 basis leads to unsightly block artifacts in the reconstruction,
which are largely avoided in the smoother and more oscillatory D4 and
D6 bases. A view of the coefficient statistics is presented in the
lowermost three panels of Fig.~\ref{wavelettopo}. The coefficients are
roughly log-normally distributed, which helps explain the success of
the thresholded reconstruction approach. While the example here was 
strictly designed to illustrate our algorithms and procedures, we
conclude that the D4 basis is a good candidate for geophysical data
representation, provided the edges between cubed-sphere chunks have
properly been accounted for.

\section{E~A~R~T~H{\hsps}M~O~D~E~L{\hsps}S~P~A~R~S~I~T~Y}

\label{tomosparse}

In tomographic studies, either as an integral part of the inversion of
after a solution has been found, the target model is parameterized by local or
global basis functions \cite[]{Nolet2008}. Blocks, cells, nodes, or voxels
\cite[e.g.,][]{Aki+77,Zhang+93,Spakman+2001,Simons+2002b,Debayle+2004,Nolet+2005a}
are all strictly local functions. Cubic B-splines
\cite[e.g.,][]{Wang+95a,Wang+98,Boschi+2004} or wavelets
\cite[e.g.,][]{Chiao+2001,Chevrot+2007,Loris+2007} are more generally
localized functions. Spherical harmonics
\cite[e.g.,][]{Dziewonski84,Woodhouse+84,Ekstrom+97,Trampert+96a,Trampert+2001}
are ideally localized spectrally but have global support
\cite[]{Freeden+99}. An intermediate approach that combines spatial
and spectral localization was developed using spherical harmonic
splines by \cite{Amirbekyan+2008a} and \cite{Amirbekyan+2008b}, but
this produces an inverse problem that scales with the square of the
number of data collected, rendering it impractical for the large-scale
tomographic systems of the future.

\begin{figure*}\centering
\iftwocol{
\includegraphics[width=0.94\columnwidth]{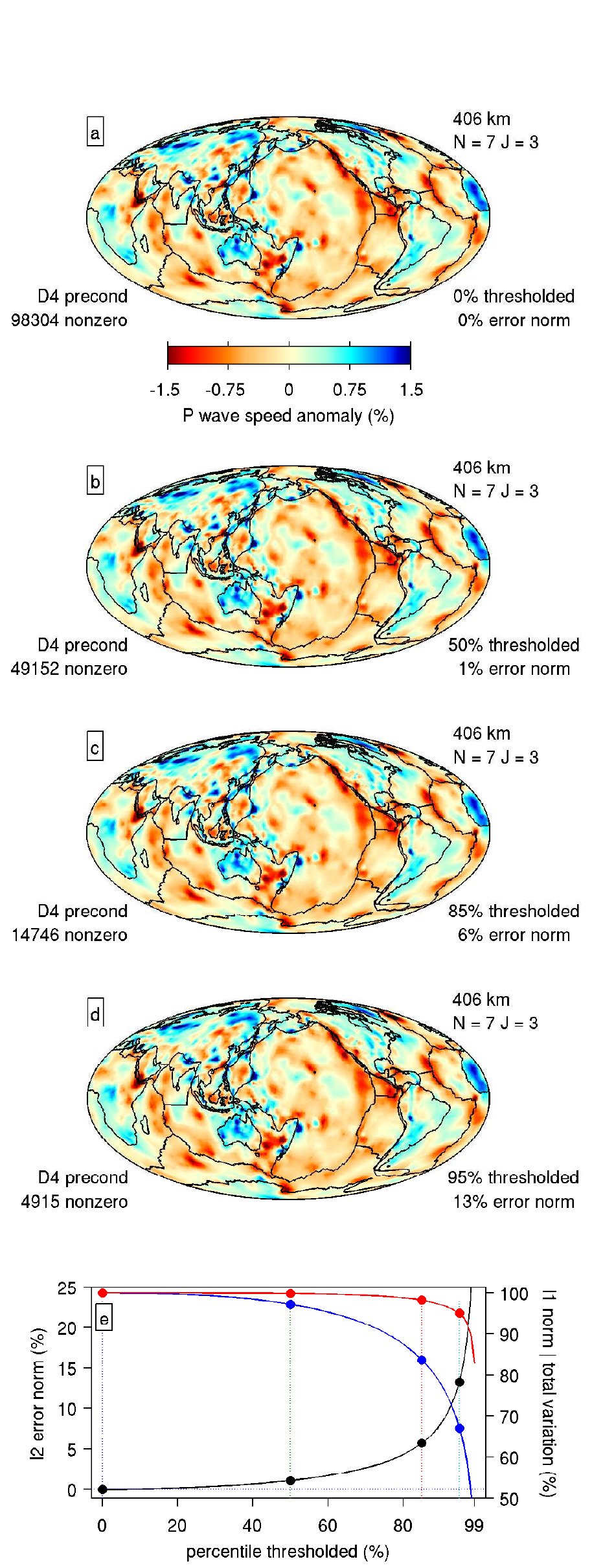}
\includegraphics[width=0.94\columnwidth]{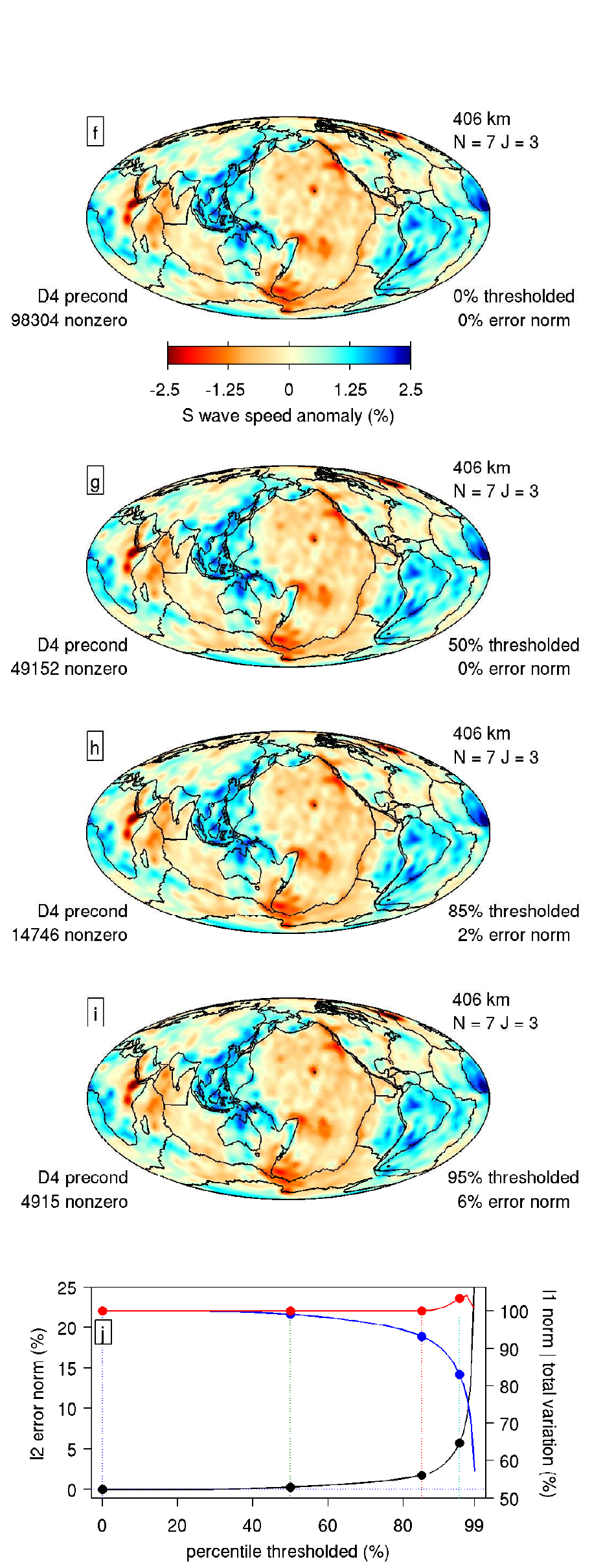}}
{\includegraphics[width=0.4\columnwidth]{loris4_GN_D4_0406}
\includegraphics[width=0.4\columnwidth]{loris4_JR_D4_0406}}
\caption{\label{waveletmontelli}\label{waveletritsema} Sparsity and
  reconstruction stability of two global seismic wavespeed models
  under incremental hard thresholding of their wavelet and scaling
  coefficients using the preconditioned edge-cognizant D4 wavelet
  basis \cite[]{Daubechies88b,Cohen+93} in the angular coordinates of
  the cubed sphere, as developed in this paper.  \textit{(a--e)}
  Results for the \textit{P}-wave seismic model
  of~\cite{Montelli+2006} and \textit{(f--j)} for the \textit{S}-wave
  seismic model of~\cite{Ritsema+2010}, at the same depth of 406~km
  below the surface of the Earth, for cubed spheres with $6\times
  2^{2N}$ elements ($N=7$), and to a $2^J$ dyadic subdivision ($J=3$).
  As a function of the percentage of the coefficients that are being
  thresholded, and relatively to the original unthresholded values,
  the bottom panels quote the spatial $\ell_2$~norms of the
  reconstruction error (in black), the total variation norms of the
  reconstructed images in the space domain (in red), and the
  $\ell_1$~norms of the coefficients that remain (in blue). The values
  obtained for the cases shown in map view are shown as filled circles
  on these graphs, and the corresponding metrics in the D2, D4 and D6
  bases are tabulated in Table~\ref{Thetable}. The reconstructions
  remain faithful to the originals even at elevated levels of
  thresholding.}
\end{figure*}

In preparing for the study of the suitability for solution of such
massive inverse problems of the wavelet transforms that we introduced
in the previous section, we take a detour in this section by
addressing the question: is the Earth sparse in a wavelet basis? Of
course we will never be able to answer this question with any degree
of certainty, but we can investigate, at the very least, whether Earth
\textit{models} are sparse in such bases. Because they are, as we
shall see, we will gain from parameterizing the inversion for future
Earth models using the spherical wavelets developed in this paper. The
expected gains are with respect to numerical efficiency but also in
terms of regularization. Since wavelets are not global functions
(ours, as can be seen from Figs.~\ref{scalesD4} and~\ref{scalesCDF},
are compactly supported, i.e. vanishing outside their scale-dependent
footprint), and yet, (bi)orthogonal, the function basis will not
dictate the model structure in areas of poor data coverage as is the
case with spherical
harmonics~\cite[]{Trampert+96b,Boschi+99,Amirbekyan+2008b}. Moreover,
though this depends on precisely what wavelet construction is being
used, they are capable of representing both smoothly varying functions
as well as preserving sharp edges, and their natural multi-resolution
nesting will allow for the model resolution to vary spatially, as
required by the data.

There is, however, another reason to find out how seismic Earth models
behave under wavelet transformation: because it enables us to study
the relative importance of model heterogeneity at different scale
lengths, which is important to help constrain geochemical and
geodynamical models and interpretations of Earth structure. The Earth
is heterogeneous at all scales but not likely everywhere to the same
degree; thermally induced deviations from the radial average
one-dimensional Earth structure are expected to be smoother and with
longer wavelengths than those due to compositional variations; the
presence of distinct scatterers further complicates this
picture~\cite[]{Shearer+2004}. In short, we are interested in
obtaining a power spectral density of sorts
\cite[][]{Chevrot+98a,Chevrot+98b,Boschi+99}, as applied to seismic
structure and how it may vary spatially within the 
Earth. As we are not in the position to return to direct measurements
of the energy distribution of heterogeneity
\cite[]{Hedlin+2000,Margerin+2003,Becker+2007,Garcia+2009} we will
instead study the sizes and scales within reported tomographic Earth models. 

\begin{table*}\centering
\begin{tabular}{rcrrrcrrr}\hline
& & 
\multicolumn{3}{c}{\cite{Montelli+2006}} & &
\multicolumn{3}{c}{\cite{Ritsema+2010}} \\
depth & thresholding & 
\multicolumn{3}{c}{relative $\ell_2$ error norm (\%)} & &
\multicolumn{3}{c}{relative $\ell_2$ error norm (\%)}\\
(km) & percentile  &
D2\hspace{0.5em} & D4\hspace{0.5em} & D6\hspace{0.5em} & &
D2\hspace{0.5em} & D4\hspace{0.5em} & D6\hspace{0.5em} \\\hline
\srule 203  & 50 &   1.816 &  0.808 &  3.025 & &  1.014 & 0.236 & 0.229\\
                & 85 &   9.212 &  4.653 &  6.324 & &  5.028 & 1.360 & 0.722\\
                & 95 &  18.721 & 11.214 & 11.456 & & 10.073 & 4.351 & 3.172\\
\srule 406  & 50 &   2.294 &  1.107 &  3.983 & &  1.267 & 0.311 & 0.297\\
                & 85 &  10.559 &  5.757 &  7.701 & &  6.182 & 1.786 & 0.968\\
                & 95 &  20.689 & 13.231 & 13.481 & & 12.393 & 5.717 & 4.125\\
\srule 609  & 50 &   2.661 &  1.244 &  3.499 & &  1.562 & 0.397 & 0.393\\
                & 85 &  11.471 &  6.419 &  7.775 & &  7.428 & 2.211 & 1.230\\
                & 95 &  21.622 & 14.145 & 14.384 & & 14.589 & 7.121 & 5.162\\ 
\srule 1015 & 50 &   3.099 &  1.311 &  3.884 & &  2.083 & 0.533 & 0.531\\
                & 85 &  12.727 &  6.896 &  8.440 & &  9.517 & 2.775 & 1.592\\
                & 95 &  23.296 & 15.107 & 15.139 & & 18.621 & 9.009 & 6.462\\
\srule 2009 & 50 &   1.995 &  0.461 &  1.799 & &  1.582 & 0.379 & 0.372\\
                & 85 &   8.890 &  3.662 &  5.174 & &  7.363 & 2.021 & 1.145\\
                & 95 &  16.946 &  9.208 &  9.104 & & 14.527 & 6.572 & 4.695\\
\hline
\end{tabular}
\caption{\label{Thetable}A companion to
Fig.~\ref{waveletmontelli}, this table lists the $\ell_2$~error norms,
relative to the original, of the reconstructions of the 
\textit{P}-wave speed model of~\cite{Montelli+2006} and the
\textit{S}-wave model of~\cite{Ritsema+2010} under hard wavelet
thresholding in the angular coordinates. See
Fig.~\ref{waveletmontelli} and text for more details. 
}
\end{table*}

From the plethora of seismic Earth models that are available to study,
we select two mantle models: one by \cite{Montelli+2006} of
compressional (\textit{P}) wavespeed heterogeneity and another by
\cite{Ritsema+2010} of shear (\textit{S}) wavespeed perturbations.
Neither model has much at all in common with the other in terms of its
construction, and from the point of view of parameterization,
Montelli's model has a tetrahedral grid underlying it, whereas
Ritsema's expands wavespeed anomalies in a spherical harmonic basis
complete to degree and order~$40$. At a depth of about 400~km,
Figs~\ref{waveletmontelli}a and~\ref{waveletritsema}f show
\textit{P}-wave \cite[]{Montelli+2006} and \textit{S}-wave
\cite[]{Ritsema+2010} anomalies from the average at that depth.
Montelli's model was \textit{interpolated} (from the tetrahedral grid
on which it was built) onto the $6\times 2^{N}$ ($N=7$) points of our
cubed sphere, whereas Ritsema's model was \textit{evaluated} (from the
listed spherical harmonic and radial spline expansion coefficients) at
these same points.  Subsequently, the wavelet transform in the D4
basis (with special boundary filters and after preconditioning, and up
until scale $J=3$) was thresholded and the results re-expanded to the
spatial grid, identically as we did for the topography in
Fig.~\ref{wavelettopo}. The results for specific values of the
thresholding (quoted as the percentile of the original wavelet
coefficients) are shown in Figs~\ref{waveletmontelli}a
and~\ref{waveletritsema}g for the 50$^\mathrm{th}$,
Figs~\ref{waveletmontelli}c and~\ref{waveletritsema}h for the
85$^\mathrm{th}$,  Figs~\ref{waveletmontelli}d
and~\ref{waveletritsema}i for the 95$^\mathrm{th}$ percentile,
respectively. At each level of thresholding the number of nonzero
wavelet/scaling expansion coefficients is quoted: at 0\% thresholding
this number is identical to the number of pixels in the surficial
cubed sphere being plotted. 

As we have written before, the wavelet transformation does not change
the \textit{number} of pieces of information with which it is
presented. Rather, it dramatically redistributes information in a
manner that allows us to simply omit those coefficients with low
values, with limited degradation to the field being represented.  This
reconstruction ``error'' can be visually assessed from the pictures;
it is also quoted next to each panel as the percentage of the root
mean squared error between the original and the reconstruction,
normalized by the root mean squared value of the original in the
original pixel representation, in percent.  Specifically, we calculate
and quote the ratio of $\ell_2$~norms in the pixel-basis model
vector~$\bm$,
\be\label{l2error}
 100\times \left.\left\|\bm-\mS
     \left\{\mT\!\left[\mW(\bm)\right]
     \right\}\right\|_2\right/ \left\|\bm\right\|_2, 
\ee 
which, in the lower-right annotations is called the ``\% error norm''.
We have written $\mW$ for any of the wavelet (analysis) transforms
that are used and $\mS$ (synthesis) for their inverses, and $\mT$ for
the ``hard'' thresholding~\cite[]{Mallat2008} of the wavelet and
scaling coefficients.

In Figs~\ref{waveletmontelli}e and~\ref{waveletmontelli}j, this same
misfit quantity~(\ref{l2error}) is represented as a black line
relevant to the left ordinate labeled ``$\ell_2$~error norm'', which
shows its behavior at 1\% intervals of thresholding; the filled black
circles correspond to the special cases shown in the map view. Only
after about 80\% of the coefficients have been thresholded does the
error rise above single-digit percentage levels, but after that, the
degradation is swift and inexorable. The blue curves in
Figs~\ref{waveletmontelli}e and~\ref{waveletmontelli}j show another
measure of misfit relevant in this context, namely the ratio of the
$\ell_1$~norms of the thresholded wavelet coefficients compared to the
original ones, in percent, or
\be\label{l1error} 100\times
\left.\left\| \mT\!\left[\mW(\bm)\right] \right\|_1\right/
\left\|\mW(\bm)\right\|_1.  
\ee
As we can see from the figure the $\ell_2$ ratios~(\ref{l2error}) in
the black curves (and the left ordinate) evolve roughly symmetrically
to the $\ell_1$ ratios~(\ref{l1error}) in the blue curves (and the
right ordinate), though evidently their range is different.

Finally, a third measure that is being plotted as the red curve is the
``total variation'' norm ratio, in percent, namely
\be\label{tverror}
100\times \left.\left\|
\bnabla\,\mS
    \left\{\mT\!\left[\mW(\bm)\right]
    \right\}
\right\|_1\right/ 
\left\| 
\bnabla\bm
\right\|_1,
\ee
whereby $\left\|\bnabla\bm\right\|_1$ is the sum over all voxels of
the length of the local gradient of~$\bm$. By this
measure, which is popular in image restoration
applications~\cite[]{Rudin+92,Dobson+96,Chambolle+97}, the quality of
the reconstruction stays very high even at very elevated levels of
thresholding; we note that its behavior is not monotonic and may
exceed~100\%. 

\begin{figure*}\centering
\includegraphics[width=0.49\textwidth]{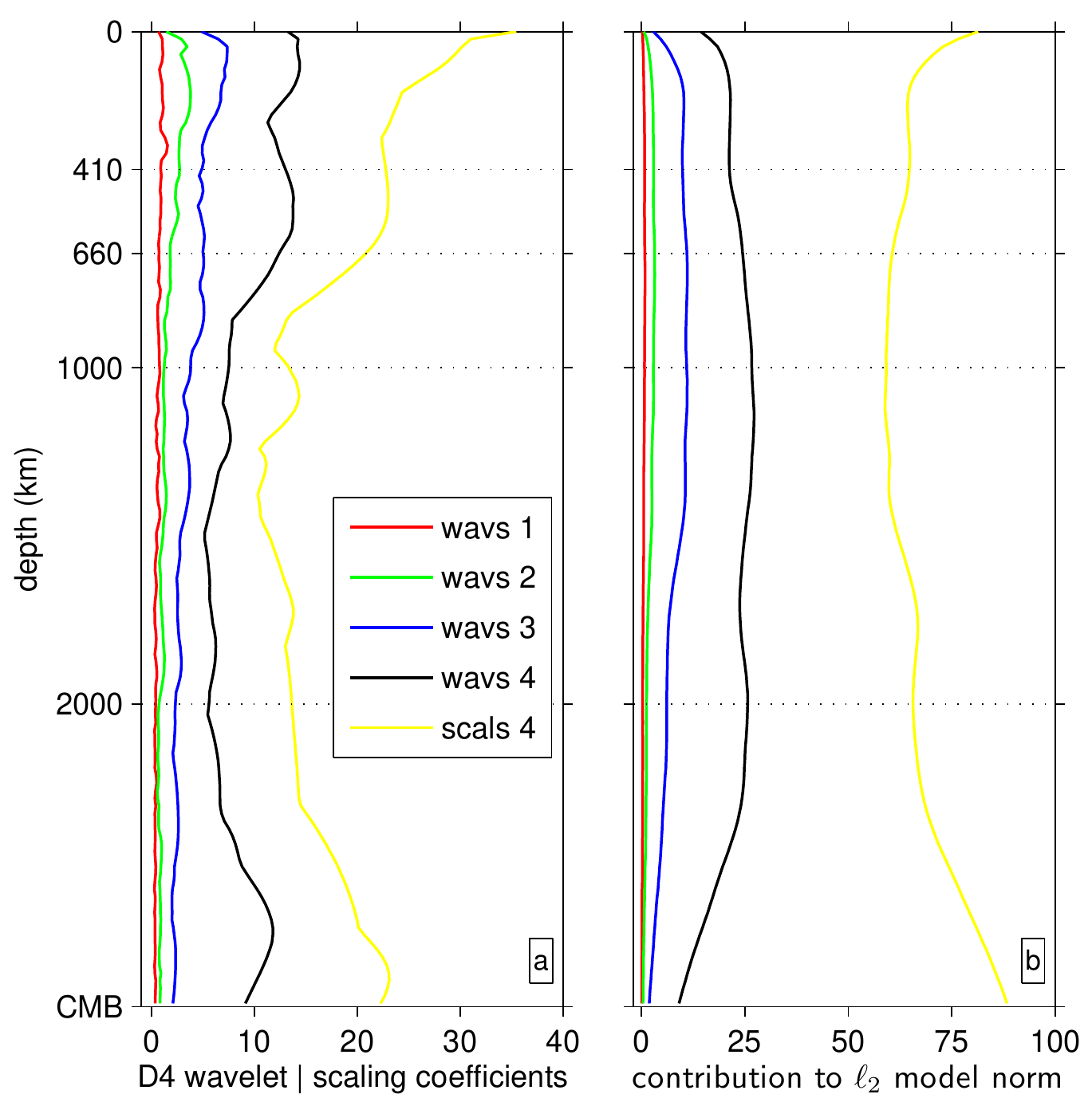}
\includegraphics[width=0.49\textwidth]{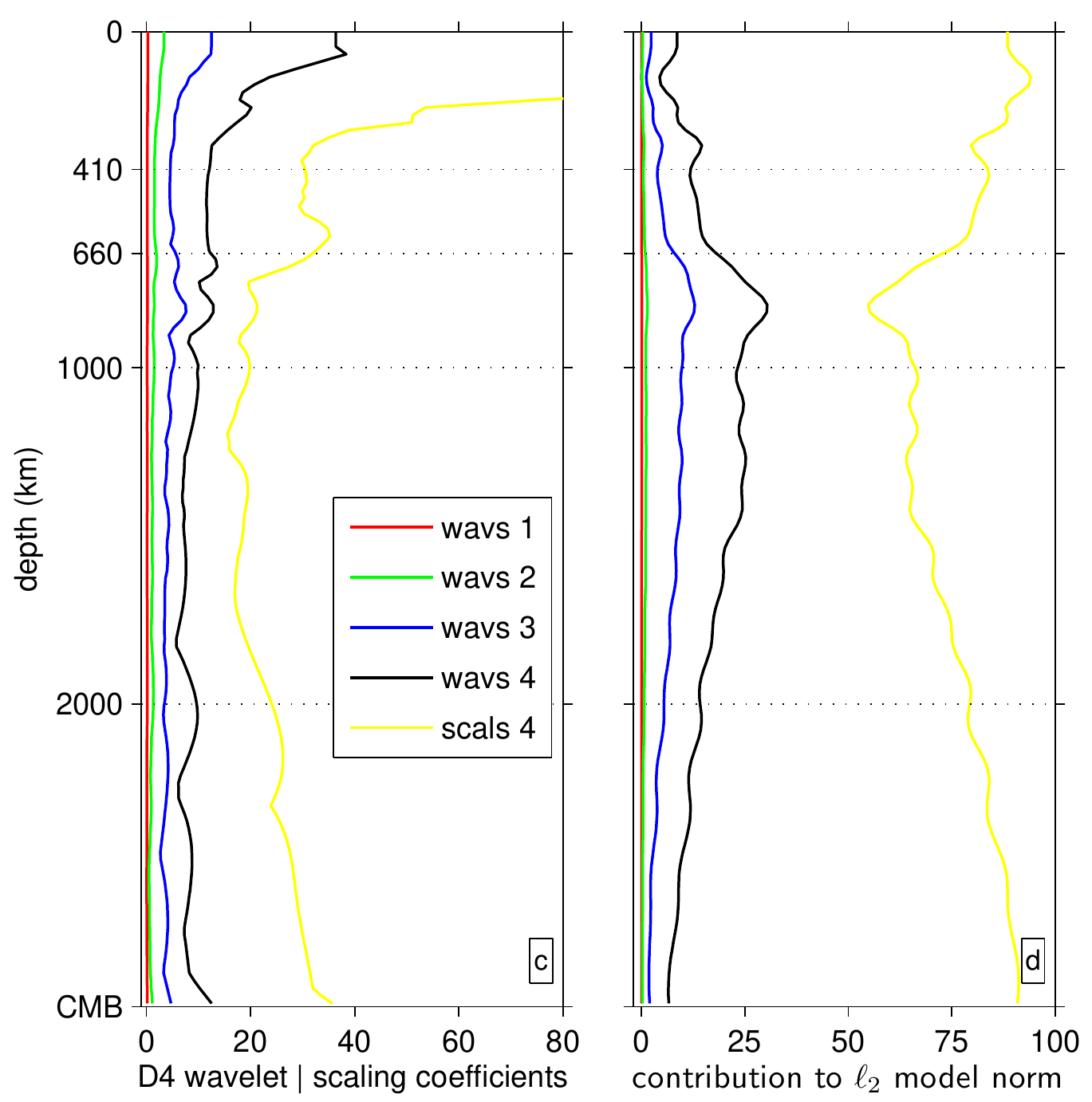} 
\caption{\label{scalemontelli}Scale lengths of seismic heterogeneity
  as a function of depth in the Earth, to the core-mantle boundary
  (CMB), obtained from the complete angular expansion in the D4
  wavelet basis of (\textit{a--b}) the \textit{P}-wave speed model of
  \cite{Montelli+2006} and (\textit{c--d}) the \textit{S}-wave speed
  model of \cite{Ritsema+2010}. See Fig.~\ref{scalesD4} for the
  wavelet and scaling functions and Fig.~\ref{waveletmontelli} for
  the seismic models: all calculations are with reference to cubed
  spheres with $6\times 2^{2N}$ elements ($N=7$), and to a $2^J$
  dyadic subdivision ($J=4$). Panels~\textit{a} and \textit{c} show
  the maximum absolute values of the wavelet or scaling coefficients
  (\textit{wavs} and \textit{scals} in the legend, respectively) at
  the scales quoted, differentiated by color. The scaling coefficients
  at the fourth scale have the largest values: at all depths the
  maximum at this scale and the overall maximum (not shown)
  coincide. Panels~\textit{b} and \textit{d} show the proportion (in
  \%) of the contribution to the overall $\ell_2$~norm of the seismic
  models at every depth by the ensemble of the coefficients at each of
  the scales. Ritsema's model has much more structure in the top
  410~km of the Earth (not shown because of the axis truncation is a
  peak with a value of 137.2 centered at 135~km) compared to the
  bottom 1000~km, as opposed to Montelli's model which has a more
  uniform distribution of heterogeneity. Both models are characterized
  by minima of seismic structure at mid-mantle depths.}
\end{figure*}

As with terrestrial topography in
Fig.~\ref{wavelettopo} we conducted all of the experiments on the
seismic models that are presented in Fig.~\ref{waveletmontelli} in the
D2, D4 and D6 wavelet bases. A summary of the $\ell_2$~error norm
ratios as a function of thresholding levels for each of those bases is
presented in Table~\ref{Thetable}. On the strength of its behavior
under the criteria~(\ref{l2error})--(\ref{tverror}) and upon visual
inspection of the results, we conclude that the D4 basis remains a
very appropriate choice for the efficient representation of seismic
models. To this choice we adhere in the geophysically motivated study
of mantle structure in those same models which follows below.   

\section{T~O~M~O~G~R~A~P~H~I~C{\hsps}M~O~D~E~L{\hsps}S~T~R~U~C~T~U~R~E}
\label{tomosparse2}


There is much geophysical interest in tying seismic observations of
mantle structure to models incorporating geodynamic modeling and
mineral physics observations
\cite[e.g.][]{Jordan+93,Karason+2000,Becker+2002,Bull+2009}. Our study
is an attempt to provide a flexible, quantitative, multiresolution
framework for such analyses that may add to the more traditional
power-spectral \cite[e.g.][]{Becker+2002,Houser+2009,Schuberth+2009}
and statistical analyses \cite[e.g.][]{Hernlund+2008}. In obliterating
the phase of the anomalies, the former line of inquiry largely loses
the relative spatial location of seismic structure, while the latter
type of study is no longer sensitive to its scale and wavelength
dependence. While in this paper we do not explicitly study the radial
correlation of mantle structure~\cite[]{Puster+95,Hilst+99a}, the
analysis below readily lends itself to adaptation in the third
dimension: our study is thus as much an initial exploration into the
richness of the wavelet transform as a way of characterizing
terrestrial heterogeneity as an encouragement to further study.

The first breakdown is as a function of depth and by scale of the D4
decomposition, as shown in Fig.~\ref{scalemontelli}. To aid in the
interpretation we remind the reader of the dominant wavelengths that
are represented at a specific scale by referring to
Fig.~\ref{scalesD4}, where of course it should be noted that the area
of the panels decreases with the square of the depth in the Earth.

The main observations relevant to both the \cite{Montelli+2006} and the
\cite{Ritsema+2010} models are that seismic wavespeed heterogeneity
has a dominantly ``red spectrum''
\cite[][]{Chevrot+98a,Chevrot+98b,Boschi+99}. Figs.~\ref{scalemontelli}a
and~\ref{scalemontelli}c show the maximum absolute values of the
wavelet and scaling coefficients at each of the four scales in the D4
decomposition, as a function of depth. The scaling functions at
scale~4 (denoted ``scals~4'' in the legend; these are depicted in
Fig.~\ref{scalesD4}a) require the largest expansion coefficients; the
maxima of the coefficients corresponding to the wavelets at scale~4
(``wavs~4'', see Figs~\ref{scalesD4}b--d) are only about half as
large; those at scale~3 (``wavs~3'', see Figs~\ref{scalesD4}e--g) peak
at about half that; and so on. \enlargethispage{1.75em}
While noting that the Montelli model
has peak amplitudes that are about half as large as the ones in the
Ritsema model, in both models the overall largest values are in the
lithosphere, which encompass  the crust and shallowmost mantle down to
about 250~km. The upper mantle (down to 660~km) and the transition
zone (410--660~km) in particular are characterized by strong maxima
that fluctuate with depth. Both seismic models have a somewhat
different take on this measure of mantle structure: the maxima in the
Ritsema model (Fig.~\ref{scalemontelli}c) are more oscillatory with
depth and have a strong peak around the 660~km mantle discontinuity
which is broader than the corresponding one in the Montelli model
(Fig.~\ref{scalemontelli}a). Each of the curves in
Figs.~\ref{scalemontelli}a and~\ref{scalemontelli}c decays sharply 
with increasing depth in the lower mantle below 660~km depth to reach
their smallest maxima in the mid-mantle before increasing again in the
bottom 1000~km, near the core-mantle boundary (CMB). This
identification of dominantly long-wavelength structure near the
core-mantle boundary~\cite[see also][]{Wysession96,Hilst+99a} is
relatively more pronounced in the 
Montelli model than in Ritsema's. In Montelli's \textit{P}~wave model 
(Fig.~\ref{scalemontelli}a) both scales~3 and~4 have significant
``bumps'' near the CMB, while the corresponding increase in maximum
structure in Ritsema's \textit{S}~wave model
(Fig.~\ref{scalemontelli}c) is more gradual and confined mostly to the
longest-wavelength scaling functions at scale~4~\cite[see
  also][]{Wysession+99}. 

\begin{figure*}\centering
\includegraphics[width=0.85\textwidth]{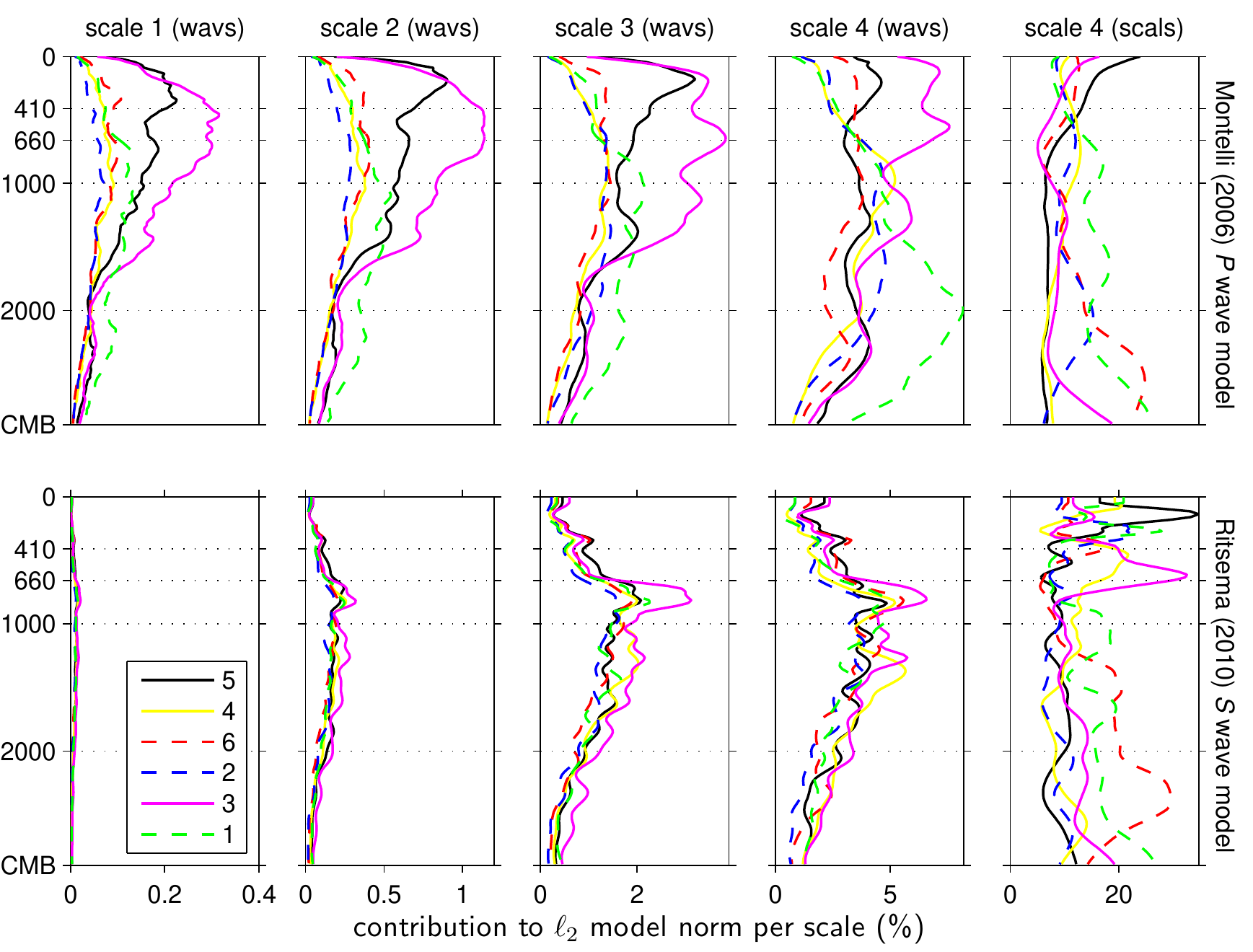}
\caption{\label{scalechunk}
Scale lengths of seismic heterogeneity as a function of depth in the
\cite{Montelli+2006} \textit{P}-wave and the \cite{Ritsema+2010}
\textit{S}-wave models. The calculations are identical to those
reported in Fig.~\ref{scalemontelli} but they are now broken per
cubed-sphere chunk to reveal geographical variations in seismic mantle
structure. See Fig.~\ref{twodtopo} for the numbering scheme used in
the legend identifying the colored lines: roughly speaking,
1~corresponds to the Pacific, 2 to Antarctica, 3 to Asia, 4 to 
South America, 5 to North America, and 6 to Africa. The relative lack
of fine structure at scales 1 and 2 and the less geographically
differentiated character at scales 3 and 4 of the Ritsema model
clearly distinguishes it statistically from the Montelli model. Other
features are more persistent between models, such as the predominantly
large-scale structure near the core-mantle boundary underneath Africa
and the Pacific, and the predominantly smaller-scale features in the
shallow mantle and crust underneath Asia and North America.}
\end{figure*}

The maximum values of the expansion coefficients in the wavelet basis
provide but one part of interpretation of mantle structure, thus in
Figs.~\ref{scalemontelli}b and~\ref{scalemontelli}d we plot the
percentage-wise relative contribution of the wavelet and scaling
coefficients at each scale to the overall $\ell_2$~norm of the
respective seismic models. These curves again reveal the scale and
depth dependence of mantle heterogeneity, but now in terms of how much
variance is explained by each scale at every depth individually: each
of the curves sums to very nearly 100\% at every depth. Their failure
to sum to \textit{exactly} 100\% arises from the preconditioning of
the wavelet transforms at the edges, which renders even the D4
transforms slightly non-orthonormal overall; however, these small
($<1\%$) deviations are not sufficiently important to influence any of
the interpretations.  In this analysis we note that once again the
relative contributions to model structure are more variable with depth
in the Ritsema model (Fig.~\ref{scalemontelli}d) than in the Montelli
model (Fig.~\ref{scalemontelli}b), which is particularly smooth in
this regard. In both, however, the importance of the structure at
scale~3 grows as a function of depth to reach a maximum about one
third of the way down. This maximum is particularly well pronounced in
Ritsema's model where it is well localized at the top of the lower
mantle, between 660~km and 1000~km depth. The growth of scale~3
structure comes at the expense of scale~4 structure, suggesting that
in that depth range long-wavelength heterogeneity is broken down to
smaller scales.

A final window into the Earth's structural heterogeneity as well as a
useful comparison between models comes in the form of
Fig.~\ref{scalechunk}, where we are able to deconstruct both of the
seismic models under consideration on a chunk-by-chunk basis. The
(arbitrary and thus easily modified) choice we made in
Fig.~\ref{twodtopo} to deviate from the canonical \cite{Ronchi+96}
orientation of the cubed sphere by approximately centering each of the
faces on a major continental landmass now allows us to study the
relative contributions of the depth-dependent seismic structure broken
down by preponderant scale length as a function of 
location in the Earth. Each of the curves originally plotted in
Figs~\ref{scalemontelli}b and~\ref{scalemontelli}d degenerates to six
individual ones with their own geographical affiliation. The
numbering scheme is the one introduced in Fig.~\ref{twodtopo}, thus in
order of appearance, 1~corresponds to the Pacific realm, 2 to
Antarctica, 3 to most of Asia, 4 to South America, 5 to North America
and parts of Eurasia, and 6 to Africa, the  middle East and the
Arabian Peninsula. In the computer code that accompanies this paper
any other wholesale rotation may be applied to the master grid,
e.g. to undo the somewhat unfortunate splitting of Australia over
chunks~2 and~3 and of Eurasia over chunks~3, 5 and~6. In other words,
the cubed-sphere wavelet transform may be applied in ``detector'' mode
by rigid rotation to center on any point of interest. Moreover,
provided the scales to be analyzed allow it, any geographical portion
of the wavelet-transformed coefficients may be zeroed out to provide
even more geographical selectivity without compromise. Such is the
power derived from multi-resolution and scale-space localization under
the wavelet transform. 

Among other features the results presented in Fig.~\ref{scalechunk}
reveal how  the dominantly long-wavelength structure near the
core-mantle-boundary is mostly due to what lies beneath Africa and the
Pacific: indeed these are regions that have been long known for being
the source of various long-wavelength mantle upwellings or
(super-)plumes~\cite[]{Ni+2003}. As to Ritsema's model, it is
surprising how little mantle structure is present at the very shortest
wavelengths of scale~2 (see Figs~\ref{scalesD4}h--i) and scale~1
(whose footprint, not shown in Fig~\ref{scalesD4}, is exactly half that of
scale~2). While also in Montelli's model the heterogeneity at these
scales remains limited at the sub-percentage level, there is
considerable more energy that contributes to the model norm, and there
is much more geographical variability between chunks in this latter
model. The relative lack of a geographical signature when comparing
Ritsema's to Montelli's model continues to be apparent at the larger
scales~3 and~4; only at scale~4 do both models ascribe mantle
structure with significant difference to each of the six gross mantle 
domains. Presumably this rather different character between both
models is due to the data selection and model parameterization:
Ritsema's model contains the effect of the whole-mantle sensitivity of
normal-mode splitting functions and the spread-out influence of
long-period surface waves. Moreover,  this model is derived in terms
of global spherical harmonics~\cite[]{Ritsema+2010}, although the
resolution gains from including spherical harmonic basis functions to
degree and order~40 as compared to an earlier iteration of this
model~\cite[]{Ritsema+99,Ritsema+2004,Ritsema2005} appear
modest. Montelli's model, in contrast, contains only body-wave
observations, albeit using finite-frequency sensitivity theory which
noticeably ``fattens'' their traditional, ray-theoretical, zone of 
influence~\cite[]{Montelli+2004b,Montelli+2006}, and it is
parameterized on a grid of tetrahedral nodes that, while globally
distributed throughout the Earth's volume, allows for more degrees of
freedom and hence spatial variability in the recovered seismic
model. Undoubtedly the scale- and space-dependent breakdown of both
models is also influenced by the different choices of
damping and smoothing in the inverse problem that led to their
construction~\cite[]{Boschi+99}. Thus, while our analysis cannot claim
to uncover the ``truth'' in characterizing Earth structure, it does
however, endow us with a measurement tool for the multi-scale
dependence of seismic \textit{model} structure. This will serve as a
target to reconcile  such models with what we can learn from forward
geodynamical modeling or in their confrontation with mineral physics
observations~\cite[e.g.][]{Megnin+97,Piromallo+2001,Cammarano+2005,Piazzoni+2007,Ritsema+2007,Bull+2009}.   


\section{T~H~E{\hsps}I~N~V~E~R~S~E{\hsps}P~R~O~B~L~E~M}
\label{inverseproblem}

In the previous sections we have constructed a new wavelet transform
on the three-dimensional ball. We have shown that, in a suitably
chosen wavelet basis, Earth models require few significant
coefficients. We have used our wavelet scheme to deconstruct two
tomographic Earth models and evaluated those both for their sparsity
and to study the distribution of mantle structure as a function of
scale, depth, and geographical location. While we have argued that we
can learn much from such exercises, we have only partially reached our
end goal, which is to harness the power and performance of spherical
wavelet bases to build \textit{new} seismic tomographic models,
directly from the data, and which are expected to be sparse in such
bases. We have not solved any inverse problems yet. In this section we
explain how these wavelets can be used to do that, too.

Wavespeed models are constructed from seismic data. With respect to
a reasonably sized global model parametrization these data are 
incomplete, as seismic stations are mostly concentrated in a
limited number of regions around the globe --- that is, until the
oceanic arrays of the future~\cite[][]{Simons+2009a}. As
usual we shall assume that a background velocity model is known, and
that our goal is to solve the data for a perturbation~$m(\bx)$ to that
reference model. We may approximate the seismic observations  
\be
\int_\oplus K(\bx)\, m(\bx)\,d^3\bx= d
,
\ee
which are of the most general kind described by such integral
equations and with~$K$ any of a veritable plethora of possible kernel
functions \cite[]{Nolet2008}, by the discretization on the grid defined in
Section~\ref{firstconstructionsect}. This leads to an inverse problem in
matrix form,
\be\label{Kmd}
\bK\cdot\bm=\bd
,
\ee
where the aim is to reconstruct the model values~$\bm$ from the data
vector~$\bd$. The elements of~$\bm$ are the values of
the model inside of each voxel and the elements of every row of~$\bK$
will be the numerical values of the integral of the kernel
$K(\bx)$ over those voxels. 

Eq.~(\ref{Kmd}) remains beholden to the usual assumption of linearity
in linking the model perturbation~$\bm$ to the data~$\bd$.
Acknowledging that the data may be contaminated by (Gaussian) 
noise~$\bn$, the inverse problem is defined as requiring us to find
the best choice of~$\bm$ by which to reduce the data misfit: the
squared $\ell_2$~norm $\|\bK\cdot\bm-\bd\|^2_2$, to the noise level,
$\|\bn\|^2_2$.  Because the data are incomplete, the problem is
ill-posed and infinitely many such models exist. Additional
conditions need to be imposed to arrive at a unique and physically
acceptable solution. This is often done by adding a penalty term
$\mP(\bm)$ to the data misfit, which leads to the functional 
\be\label{penalizedfunctional}
\mF(\bm)=\|\bK\cdot\bm-\bd\|^2_2+\mP(\bm)
,
\ee
which is to be minimized. The role of the penalty term is to ensure
that~$\mF$ has a unique and acceptable minimizer. The trade-off
between data fit and \textit{a priori} information is encoded in the
penalty~$\mP$. A convenient and often advocated choice for~$\mP(\bm)$
is a multiple of the norm-squared of the Laplacian of
the model, $\mP(\bm)=\lambda \|\nabla^2 \bm\|^2_2$, which favors
smoothness in the solutions~\cite[see,
e.g.,][]{Yanovskaya+90,VandeCar+94}. The equations for the
minimum of~$\mF(\bm)$  remain linear, 
\be
\bK\Trm\cdot\bK\cdot\bm+\lambda (\nabla^2)\Trm\nabla^2 \bm=\bK\Trm\cdot\bd,
\ee
and can thus be handled by standard algorithms. The trade-off 
parameter~$\lambda$ needs to be carefully chosen \cite[]{Hansen92}.

The novelty now is that we should be able to use model
\textit{sparsity} rather than smoothness as prior information. As
discussed in  Section~\ref{tomosparse} seismic tomographic models may
be very well represented by a sparse wavelet expansion. Incorporating
this knowledge from the start may therefore lead to important benefits
to the behavior of the inversion scheme. In the following we shall
 assume that we have chosen a particular set of wavelet and scaling
basis functions, see Section~\ref{firstconstructionsect}, to
represent and build the unknown model.  

With the model~$m(\bx)$ expanded in our wavelet basis via the
transform~$\mW$ as in the notation of
Section~\ref{tomosparse}, and the individual basis functions collected 
in the columns of a matrix~$\bS$, the synthesis map, the
pixel-basis model vector~$\bm$ is 
\be\label{mSw}
\bm=\mS(\bw)=\bS\cdot\bw
,
\ee
with~$\bw$  the vector of expansion coefficients 
in this basis. 
In having previously defined our construction in
terms of a discrete wavelet transform we do not need to devise a
separate form of discretization for each of the many choices of
wavelet bases that are available to us. In this flexible approach we
define the grid size of the cubed sphere at the outset and  we are
thus able to switch between the various wavelet bases without much
additional effort.  As we shall remark later on $\bS$ will usually be
provided as a (fast) software algorithm and not as a matrix \textit{per 
se}. We shall also see that the seismic inversions only require
application of $\bS$ and its transpose~$\bS\Trm$.  The
inverse~$\bS^{-1}$, the analysis map, is not required to be known ---
or even exist, as is the case for a redundant set of basis
functions.  

The sparsity of the model parameters~$\bw$ can now be encouraged by
choosing the penalty~$\mP$ to be proportional to the number of nonzero
entries in $\bw$, which we write as $\|\bw\|_0$ for short. The
functional to be minimized then becomes 
\be
\mF_0(\bw)=\|\bK\cdot\bm-\bd\|^2_2+\lambda \|\bw\|_0
.\label{l0functional1}
\ee
We define the solution to the inverse problem as
\be
\bwh=\arg\min_\bw\left(\|\bK\cdot\bS\cdot\bw-\bd\|^2_2+\lambda
\|\bw\|_0\right)
,\label{l0functional2}
\ee
and the reconstructed model is
\be\label{recon}
\bmh=\bS\cdot\bwh
.
\ee The functional
in eq.~(\ref{l0functional2}) however is not convex: there exist local
minima which makes the minimization much less feasible than solving a
system of linear equations. Despite this an iterative algorithm based
on hard thresholding exists
\cite[]{Blumensath.Davies2008,Blumensath.Davies2009a}, as briefly
discussed by \cite{Loris+2010}. 

An alternative, and computationally much more tractable, method
for imposing model sparsity in a given basis is to use an
$\ell_1$~norm penalty
\cite[]{Donoho2006,Daubechies+2004,Bruckstein+2009}.
By identifying $\|\bw\|_1=\sum_i |w_i|$, and choosing
$\mP=2\lambda\|\bw\|_1$ for the penalty function, then
\be\label{l1functional2} 
\mF_1(\bw)=\|\bK\cdot\bS\cdot\bw-\bd\|^2_2+2\lambda
\|\bw\|_1
\ee
is to be minimized. This functional is convex: a local minimum is
therefore automatically a global minimum \cite[]{Loris+2007}. 

The functional (\ref{l1functional2}) is not differentiable but
because $\mF_1$ is the sum of a differentiable and a separable
non-differentiable part, convex optimization techniques can  find
$\bwh=\arg\min_{\bw} \mF(\bw)$ and the corresponding
model~(\ref{recon}) with reasonable efficiency. Indeed the iteration 
\begin{subequations}
\label{algo}
\be
\bw_{n+1}=\mU\!\left[\bw_n+
\beta_n(\bw_n-\bw_{n-1})\right],
\label{fista}
\ee
\be
\mU(\bw)=\mT_{\alpha\lambda}\!\left[\bw+\alpha\,\bS\Trm \cdot\bK\Trm 
\cdot(\bd-\bK\cdot\bS\cdot\bw)\right]\label{fista2}
,\ee
converges to the minimizer of~(\ref{l1functional2}), as shown by
\cite{Beck+2008}. Hereby~$\mT_{\alpha\lambda}$  now stands for ``soft''
thresholding~\cite[]{Mallat2008} of the coefficients on a
component-by-component basis, which is to say $\mT_\tau(\bw)=0$ for
$|\bw|\leq \tau$, and $\mT_\tau(\bw)=\bw-\tau\,\mathrm{sgn}(\bw)$ for
$|\bw|>\tau$.  This is a non-linear operation. The parameter $\alpha$
in eq.~(\ref{fista2}) can be chosen as the reciprocal of the largest
eigenvalue of $\bS\Trm\cdot\bK\Trm\cdot\bK\cdot\bS$. We choose $t_0=1$
and
\ber
\beta_n&=&(t_n-1)/t_{n+1},\\
t_{n+1}&=&(1+\sqrt{1+4t_n^2})/2\label{fista3}
.
\eer
\end{subequations}
A non-iterative direct algorithm also exists
\citep[]{Efron+2004,Loris2008}, but because of the large problem sizes
typically encountered in seismic tomography, we focus here on this
so-called Fast Iterative Soft Thresholding Algorithm (FISTA). It has
an $1/n^2$ rate of convergence to $\mF_1(\bwh)$, which is in a sense
optimal. The algorithm~(\ref{algo}) was used by \cite{Loris+2010} on a
3D toy tomographic model. There is however a typographical error in
that work, which missed the factor~$4$ under the square root in
eq.~(\ref{fista3}).

\begin{figure}\centering
\includegraphics[width=0.75\columnwidth]{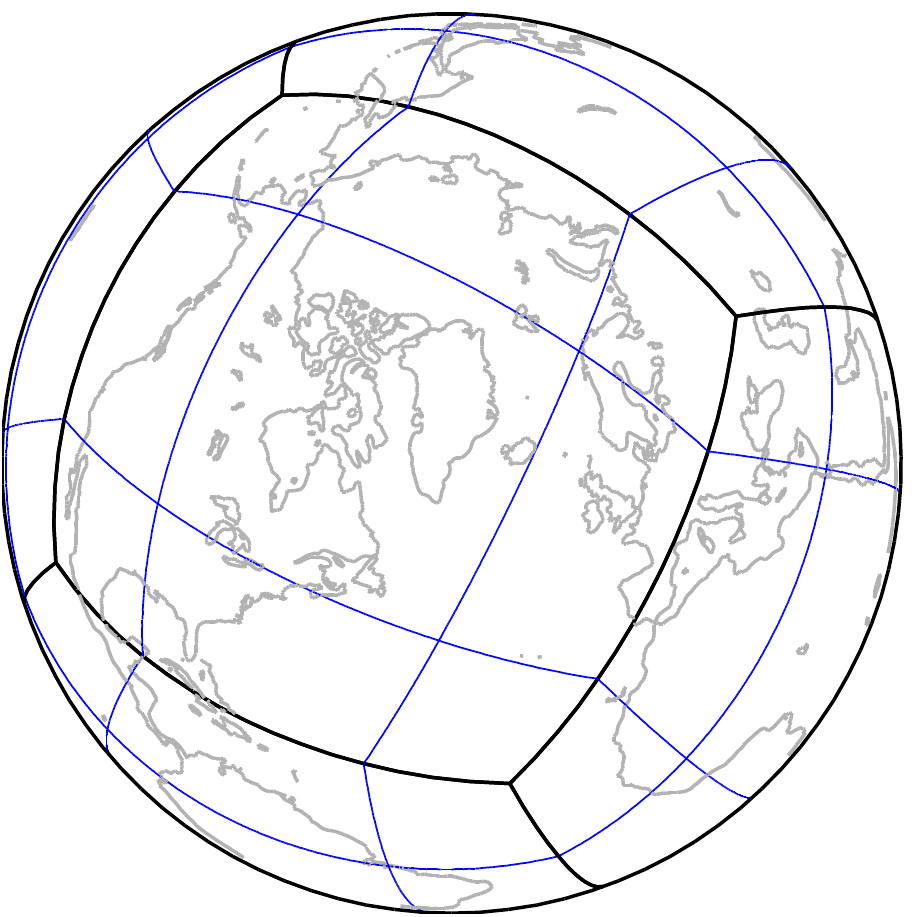}
\caption{\label{thesuperchunk}Aerial view showing our second adaptation of
the cubed sphere of \cite{Ronchi+96}. The black lines identify the
boundaries of the six chunks that were apparent also in
Fig.~\ref{thechunk}. The blue lines correspond to the boundaries of
the overlapping ``superchunks'' as discussed in the text.}
\end{figure}

The iterative algorithm~(\ref{algo}) requires only two linear maps,
and their transposes. First there is the linear map from model to data
space, given by the matrix~$\bK$ in eq.~(\ref{Kmd}).  The second is
the linear map~$\bS$ from model parameters to model space,
eq.~(\ref{mSw}). This map is typically available in the form of a
(fast) algorithm, \textit{in casu} the inverse wavelet
transform~$\mS$, rather than explicitly in matrix form.
Each iteration step of algorithm~(\ref{algo}) requires one application
of $\bK$, $\bK\Trm$, $\bS$ and $\bS\Trm$ each.
Eqs~(\ref{fista}--\ref{fista3}) demonstrate that the iterative
inversion algorithm does not require the inverse of the map~$\bS$,
much as it does not require the inverse of~$\bK$. Moreover, neither
$\bS$ nor $\bK$ need be invertible. As already mentioned, this means
that a model may be represented by a sparse superposition of a
redundant set of functions in which the expansion of the model is no
longer unique.  For example, redundant dual-tree wavelets were used in
a synthetic tomography experiment by \cite{Loris+2007}.

In practice it turns out to be easier to keep~$\bK\cdot\bS$ and
$\bS\Trm\cdot\bK\Trm$ in eq.~(\ref{fista2}) in factorized
form. One can then easily switch bases by modifying $\bS$ and
rerunning the inversion algorithm. No new matrix $\bK\cdot\bS$ needs
to be pre-computed, which is important given that $\bK$ may have
several hundreds of thousands of rows. This is particularly useful in
the case of sparse reconstructions, where the choice of basis itself
is one of the undetermined factors to be assessed by simulation during
the inversion: Section~\ref{tomosparse} made it clear that using model
sparsity as \textit{a priori} information depends on the details of
the basis used. Setting up the inversion software in this 
manner is therefore forward-looking as new transforms can
easily be incorporated later. Examples of emerging techniques that can
be evaluated in this context are curvelets and shearlets
\cite[]{Candes+2005b,Labate+2005,Easley+2008},
which offer better directional sensitivity than classical
wavelet transforms but are redundant. The described flexibility
of our approach was a major design requirement and will yield many
dividends in future applications.

\section{A{\hsps}S~E~C~O~N~D{\hsps}C~O~N~S~T~R~U~C~T~I~O~N}

\label{secondconstructionsect}

In principle we are now ready to apply the first family of wavelet
constructions on the cubed sphere that we introduced in
Section~\ref{firstconstructionsect} to the inverse problem in the
manner outlined in Section~\ref{inverseproblem}. As shown in
Section~\ref{tomosparse} we expect our solutions to be sparse, and
as discussed in Section~\ref{tomosparse2} we will be able to
use this sparsity and the location- and scale-dependence of the
results to make geophysical inference about the structure of seismic
heterogeneity in the Earth.

As we recall, our First Construction entailed defining wavelet and
scaling functions on a single chunk $\xi,\eta \in [-\pi/4,\pi/4]$ and
then mapping them onto the sphere using eq.~(\ref{xieta2xyz}).  By
this definition the basis functions live on a single chunk. Without
the modifications and preconditioning of the basis at the boundaries
between the chunks that we introduced, sharp discontinuities in the
behavior of the coefficients occur at the chunk edges; making the
transforms edge-cognizant, as we did in the manner of~\cite{Cohen+93},
required special tailoring of the transforms. This is often cumbersome
and in general harms our stated goal of keeping our procedure flexible
enough to be able to switch from one wavelet family to another which
might be more suitable with hindsight. In addition, the interval
wavelet transforms that we used so far are not norm-preserving.
Extensive experimentation with such bases revealed that despite their
qualities in the representation of geophysical functions, i.e. in
performing the forward mappings, when used for the inverse problem the
solutions obtained using eq.~(\ref{algo}) were plagued by unsightly
artifacts at the seams between chunks (not shown). Presumably the
$\ell_1$-thresholding could be adapted locally to counter this effect,
but to be truly practical we should not have to resort to this. We
thus desire a mechanism to map any localized basis function defined on
a Cartesian grid to the sphere, with smoothness even across chunk
boundaries. Here we present a straightforward, universal method that
accomplishes this.

\begin{figure}
\centering
\includegraphics[width=0.85\columnwidth]{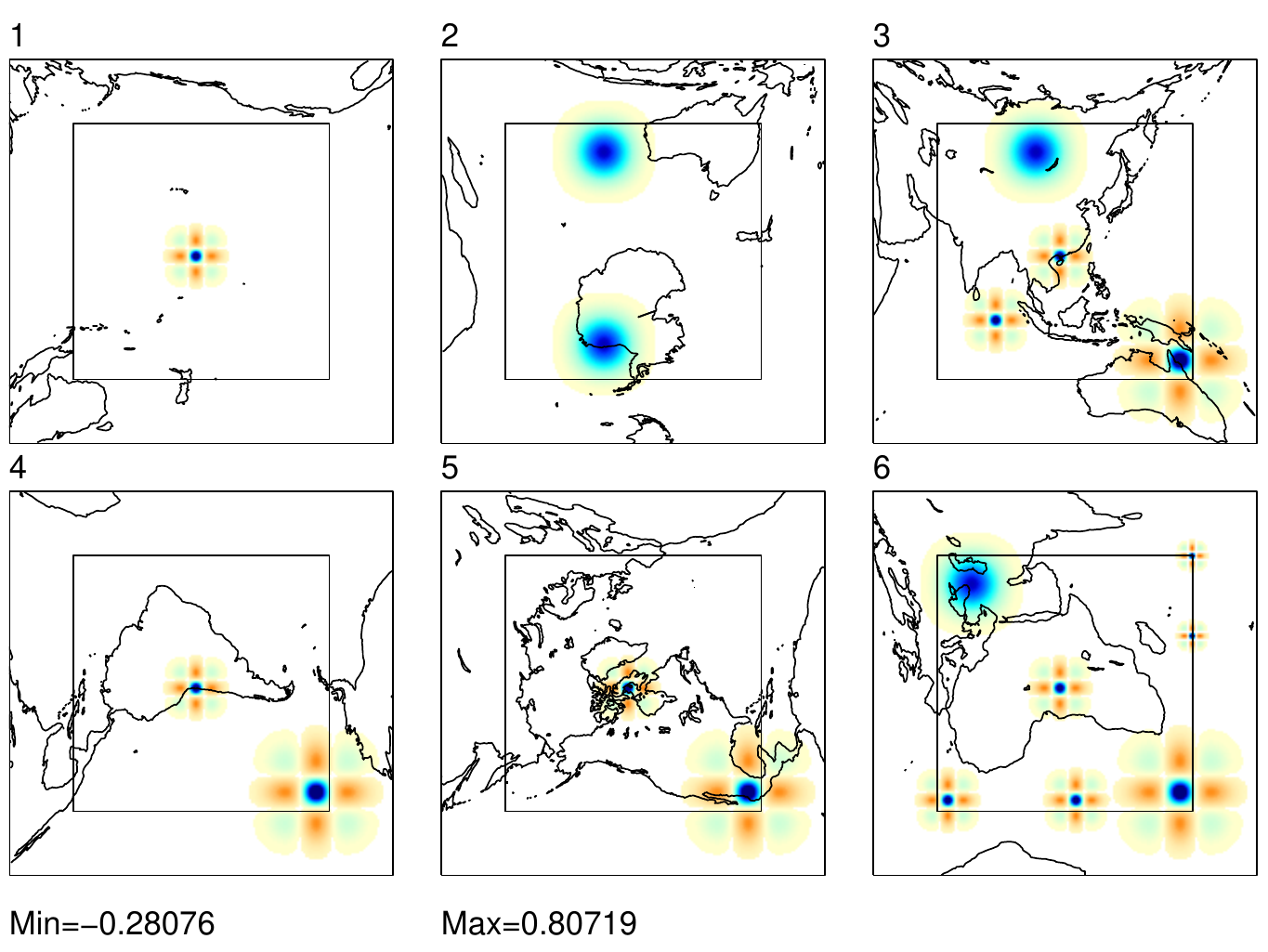}\\[2em]
\includegraphics[width=0.85\columnwidth]{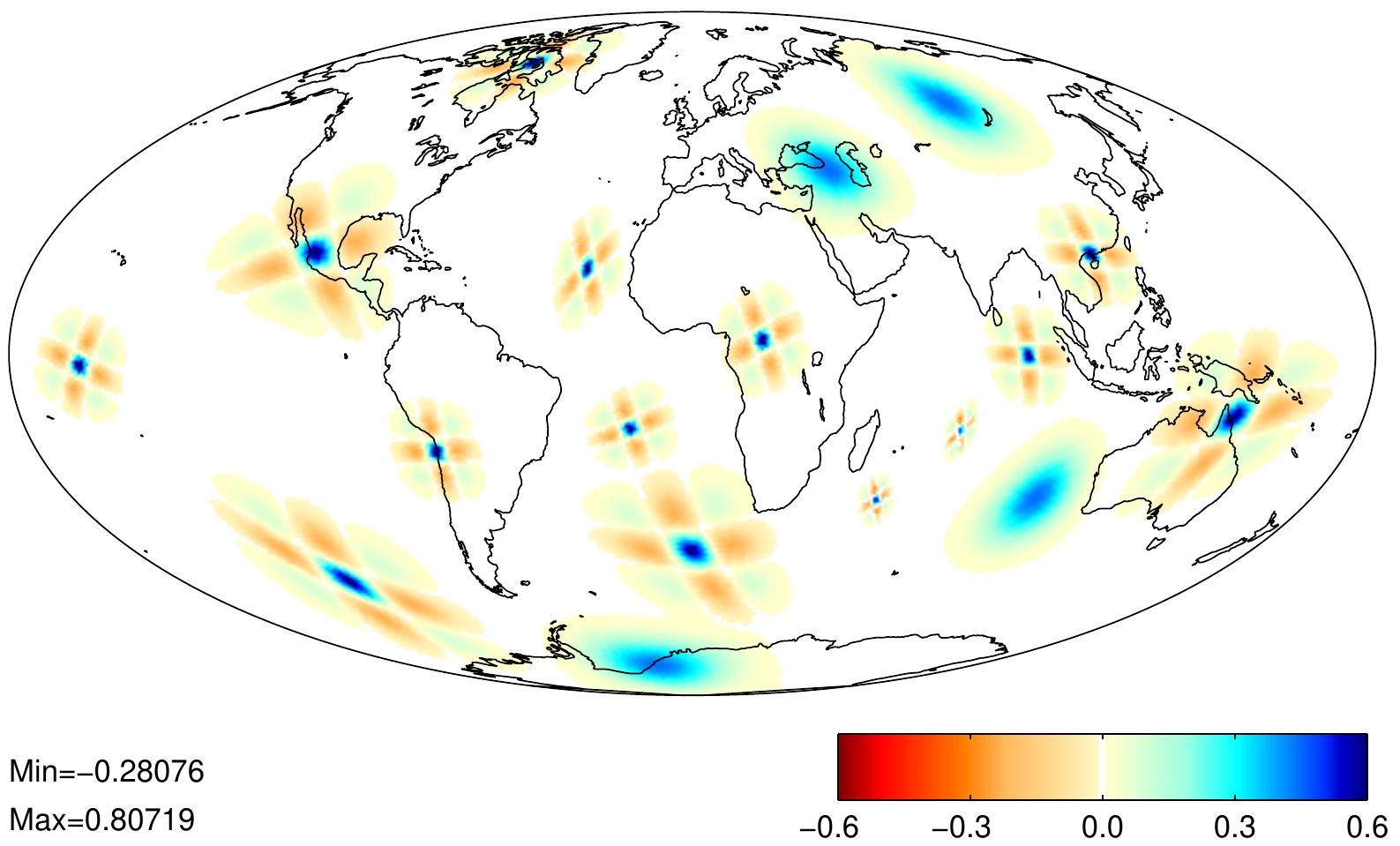}
\caption{\label{superchunksfig}Top: Six superchunks with a number of
  wavelet and scaling functions defined on them. Bottom: The same
  functions mapped to the sphere by the procedure described in
  Section~\ref{secondconstructionsect}. They are smooth everywhere.}
\end{figure}

As opposed to the geometry of the \cite{Ronchi+96} cubed sphere shown
in Fig.~\ref{thechunk}, we now cover the sphere with six larger
chunks, by extending the coordinates by $50\%$ on each chunk, to
$\tilde{\xi},\tilde{\eta} \in [-3\pi/8, 3\pi/8]$, see
Fig.~\ref{thesuperchunk}. We shall refer to these partially
overlapping domains as ``superchunks''. In $(\tilde{\xi},
\tilde{\eta})$ coordinates they are simply six large squares (rather:
cubes if we take the radial direction into account also), with the
``original'' chunks at their centers. Functions defined on this
central part can now smoothly cross into the outer part, that is, they
are allowed to spill over into another chunk while staying in the same
superchunk. Fig.~\ref{superchunksfig} shows a selection of examples
where this is the case. The smoothness of the functions across the
boundaries is apparent, though we note that if we were to plot them in
the manner in which Fig.~\ref{twodtopo} was presented, they would
appear to have kinks in them; this is simply because the coordinate
transform of eq.~(\ref{xieta2xyz}) itself is non-smooth.


To map a function defined on a single superchunk $\tilde\kappa=
1\rightarrow 6$ to the corresponding chunk and its neighbors, one
loops over all the voxels in this central chunk and its four
neighbors. The center $(\xi,\eta,r,\kappa)$ of each such voxel is
mapped to $(x,y,z)$ coordinates using formula (\ref{xieta2xyz}). In
the same way as eq.~(\ref{xyz2xieta}) we then calculate  
\be
(\tilde\xi,\tilde\eta)=\left\{\!\!
\begin{array}{lcl}
{}[\,\atan(z/y), \atan(-x/y)\,] & \mathrm{if} & \tilde\kappa=1, \\
{}[\,\atan(y/x), \atan(-z/x)\,] & \mathrm{if} & \tilde\kappa=2, \\
{}[\,\atan(-y/z), \atan(x/z)\,] & \mathrm{if} & \tilde\kappa=3, \\
{}[\,\atan(x/z), \atan(-y/z)\,] & \mathrm{if} & \tilde\kappa=4, \\
{}[\,\atan(-z/x), \atan(y/x)\,] & \mathrm{if} & \tilde\kappa=5, \\
{}[\,\atan(-x/y), \atan(z/y)\,] & \mathrm{if} & \tilde\kappa=6, \\
\end{array}
\right.
\label{xyz2xietakappafixed}
\ee
to convert these $(x,y,z)$ to the 
$(\tilde\xi,\tilde\eta,\tilde r=r)$ coordinates in  
the superchunk $\tilde\kappa$, limited to
$-3\pi/8\leq\tilde\xi,\tilde\eta\leq3\pi/8$. This then determines which
voxel in the superchunk is mapped to the  voxel in the original
chunk. The index of the voxel in each superchunk~$\tilde\kappa$ is
\be
i=1+\left\lfloor\left(\tilde\xi+\frac{3\pi}{8}\right)\frac{2N}{\pi}\right\rfloor
,\quad
j=1+\left\lfloor\left(\tilde\eta+\frac{3\pi}{8}\right)\frac{2N}{\pi}\right\rfloor
,\label{eq1}
\ee
where~$N$ is the number of voxels in the $\xi$ and $\eta$
directions of a chunk and $\lfloor\,\rfloor$ indicates rounding
down. Voxel indices run from $1\rightarrow N$ in an 
original chunk and from $1 \rightarrow 3N/2$ in a superchunk. The
central part of a superchunk is a copy of the original chunk, whereas
the voxels outside the center of a superchunk are mapped to
neighboring chunks. 
As the superchunks partially
overlap, a chunk voxel on the sphere may receive contributions from up
to three superchunks: a voxel near a chunk corner may receive three
contributions, a voxel near a chunk edge may receive two, and voxels
near chunk centers only one. The identifications are most easily made
by table look-up. 


In Fig.~\ref{superchunksfig} we show a number of wavelet
functions from this Second Construction at a variety of
locations. These now map smoothly to the 
sphere. The wavelets shown here are from the \cite{Cohen+92} CDF 
4--2 wavelet family, as in Figure~\ref{scalesCDF}. 
These are mirror-symmetric in the $(\xi,\eta)$ domain, but they are no longer
orthogonal. As in Section~\ref{firstconstructionsect} the wavelets at
a fixed scale are not rotations of each other on the sphere, but
rather translates in the superchunk $(\tilde\xi,\tilde{\eta})$
domain. This effect is most noticeable for the wavelet and scaling
functions that are located on or near chunk edges, specifically near
the corners. Basis functions that have the same norm in the superchunk
domain may not have the same norm in the chunk domain. 

\section{N~U~M~E~R~I~C~A~L{\hsps}E~X~P~E~R~I~M~E~N~T~S}

\label{experiment}

\begin{figure}
\centering
\iftwocol{
\includegraphics[width=1\columnwidth]{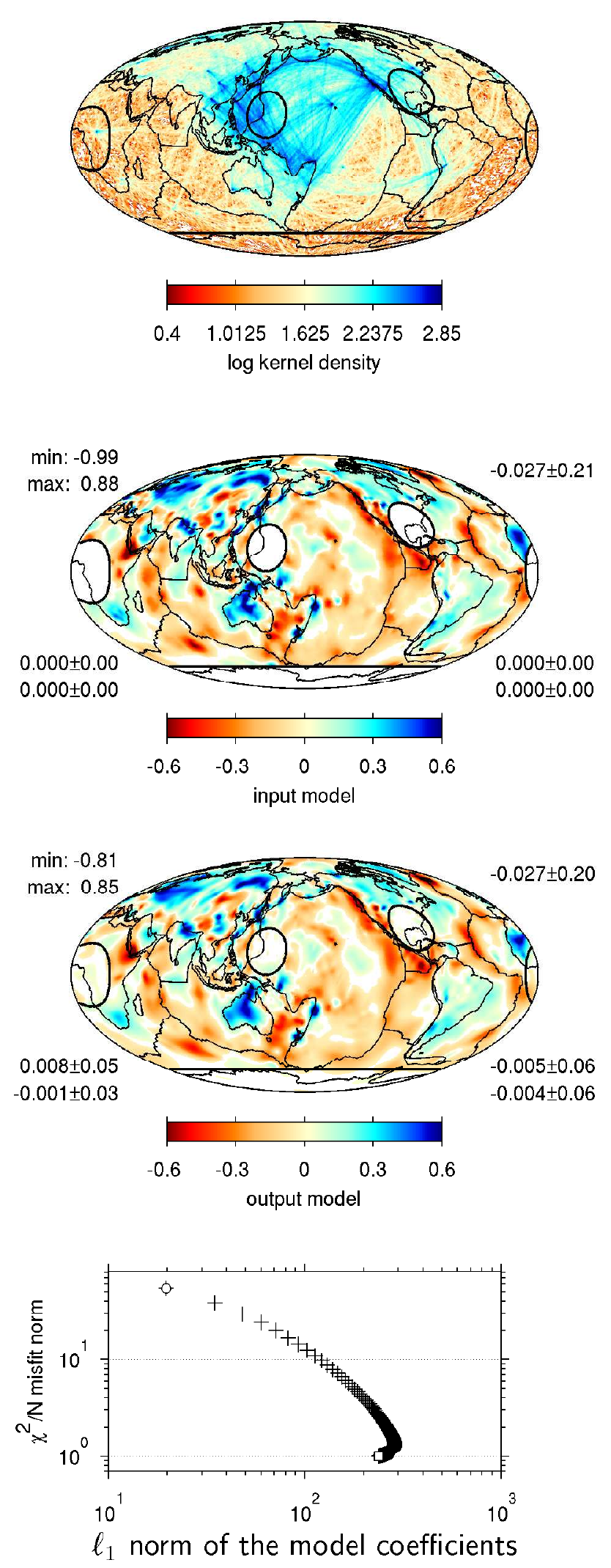}}
{\includegraphics[width=0.5\columnwidth]{loris8}}
\caption{Synthetic experiment under realistic conditions with a twist,
illustrating the recovery of a seismic tomographic model with
artificially introduced ``blank spots'' from noisy data, using the
Second Construction discussed in
Section~\ref{secondconstructionsect} and the iterative algorithm of
Section~\ref{inverseproblem}. The solution after one iteration is
represented by the filled white circle in the bottom panel; the
solution after 1000 iterations by the filled white square.} 
\label{reconstructionfig}
\end{figure}

We consider a set of great-circle paths that is a global collection of
2469 earthquakes and 199 stations yielding 8490 surface-wave paths, a
situation based on, if not identical to, the ray path coverage in the
models of Rayleigh-wave phase speeds at 80~s period obtained by
\cite{Trampert+95,Trampert+96a,Trampert+2001}. For simplicity we
convert this path coverage to the ray-theoretical values of
arrival-time sensitivity expressed in our model domain. The image in
Fig.~\ref{reconstructionfig} (top panel) renders all rays
in this data set of realistically heterogeneous global seismic
sensitivity. For synthetic input model we chose a single  interval of
the \cite{Montelli+2006} model centered on $722$~km depth, shown in 
Fig.~\ref{reconstructionfig} (second from the top). In addition, and
this is an admitted departure from realism, we select four circular
regions of null structure. Their purpose is to test the inversion
algorithm and the choice of basis when sharp wavespeed contrasts are
known to be present in the true model. We calculate the travel-time
perturbations over these $8490$~ray paths and add Gaussian noise to
them with an rms value that is $10\%$ of that of the rms of the
data. The variance of this noise is denoted~$\sigma^2$.

The reconstruction is by the algorithm~(\ref{algo}) using the
four-level CDF 4--2 wavelets under the Second Construction by which smooth
chunk crossings were enabled, as shown in
Fig.~\ref{superchunksfig}. In keeping with the 
description of Section~\ref{inverseproblem} the dual aim is to satisfy
the noisy data in the traditional $\ell_2$~sense while favoring a
model that is sparse in the wavelet basis by minimizing the
$\ell_1$~norm of the coefficients. Fig.~\ref{reconstructionfig} (third
from the top) shows the obtained solution. Of the $6\times128^2=98304$
degrees of freedom in this parameterization the algorithm terminates
on a model with $1670$ nonzeroes. Due to lack of data, the relative
output error is high: $33.5\%$.

The behavior of the solution through the 1000 iterations is shown in
Fig.~\ref{reconstructionfig} (bottom panel), which plots the
$\ell_1$~norm of the wavelet coefficients against a measure of the
evolving misfit calculated as the ``reduced chi-square''
\be\label{chitwo}
\frac{\chi^2}{N}=\frac{\|\bd-\bdh\|^2_2}{\sigma^2\|\bd\|_0}
=\frac{\|\bd-\bK\cdot\bmh\|^2_2}{\sigma^2\|\bd\|_0}
,
\ee
in other words, the squared $\ell_2$~norm of the data misfit
normalized by the noise variance and the total number of data
constraints, for which it is reasonable to assume~\cite[]{Nolet2008}
that it is distributed as a $\chi^2_1$ variable. The starting point of
the iteration is marked by the filled white circle, and the final
solution by the filled white square, which is arrived at when the
$\chi^2/N$ variable in eq.~(\ref{chitwo}) reaches its
expectation~1. Every one of the 1000 solutions in the sequence is
marked by a black cross. As we notice the algorithm~(\ref{algo})
rapidly reduces the data misfit in the first few steps, slowing down
after that, at the same time increasing the sum of the absolute values
of the wavelet coefficients. After ``turning a corner'' in this space,
the remainder of the time spent is in reducing the $\ell_1$~norm of
the coefficients of the solution while slowly converging to the target
reduced chi-square of $\chi^2/N=1$. 

The solution is very good; the input model is well matched and the
leakage of the solution into the areas where no structure should be
recovered is relatively minor. The global map views in
Fig.~(\ref{reconstructionfig}) had all values that fall below a
threshold of 1/20$^\mathrm{th}$ of their maximum absolute value
rendered white for visual guidance, and they receive several
additional annotations for us to be able to judge the quality of the
solution quantitatively. The minimum and maximum values of the models
are quoted in the top left corner, and in the top right corner we show
their mean and the root mean squared values. The four sets of numbers
in the bottom left and right hand corners quote these same metrics for
the areas contained inside of the circular areas. As the input model
has no structure there, all of these are zeroes. This is no longer the
case for the output which serves as our way to evaluate the leakage of
the solution into those areas. As we can see the comparison is very
favorable. 

While we have conducted numerous synthetic tests with a
multitude of synthetic input models (including checkerboard tests,
Gaussian shaped anomalies positioned at various locations, and using a
variety of ray path coverages), only one of these tests is reported
here. As noted by \cite{Loris+2007,Loris+2010} there are more
algorithms available to us than the one described in eq.~(\ref{algo}),
and as we have argued in this paper there is a plethora of wavelet
constructions that can be brought to bear on the inverse problem of
global seismic tomography. All of these alternatives remain in
principle candidates to be implemented using our First or Second
Construction for wavelets on the sphere. A more detailed comparison of
their relative performance is well underway and will be reported in
forthcoming work. It is there also that we will fully integrate the
third dimension into our formalism. Conceptually, there is no
difficulty in doing this: we have transformed the ball of the Earth
into six independent or partially overlapping Cartesian model domains
with three separable coordinates. Taking into account the depth
dimension merely involves applying a third wavelet transform to the
result of the transform in the two angular coordinates, but as there
remain choices to be made, a thorough discussion remains outside the
scope of the current paper.


\section{C~O~N~C~L~U~S~I~O~N~S}
\label{conclusions}

Until now, seismic wavespeed models of the Earth have been routinely
parameterized in terms of spherical harmonics, networks of tetrahedral
nodes, rectangular voxels, or spherical splines. However, there were
few approaches to Earth model parameterization by wavelets on 
the three-dimensional ball. To the rich field of wavelets on the ball
or its surface, the sphere, we have contributed two new flexible
constructions that are eminently suited to solve seismological
tomographic inverse problems.  

To form the numerical grid we considered a surface tesselation known
as the ``cubed sphere'', popular in fluid dynamics and computational
seismology, which can be combined with a (semi-regular) radial
subdivision. This mapping transforms the entire volume of the mantle
into six portions. In the new variables, these ``chunks'' correspond
to rectangular boxes with Cartesian coordinates. Standard algorithms
can then be used to perform the wavelet transformation (or any other)
in each of the six bounded volumes. We developed two possible classes
of discrete wavelet transforms in the angular dimension of the cubed
sphere. One relies on preconditioning and special boundary filters to
account for the edges separating the chunks; another one broadens the
definition of the cubed sphere to include chunks that partially
overlap, on which we implement standard wavelet transforms.

Much has been gained by our design of procedures that efficiently
parameterize the seismological inverse problem. First, the
multiresolution character of a wavelet basis allows for the models to
be represented with an effective spatial resolution that varies as a
function of position within the Earth. Second, inversion schemes that
are formulated in terms of wavelets can exploit recent theoretical and
numerical advances by which the most sparse solution vector, in 
wavelet space, is found through iterative minimization of a
combination of the $\ell_2$ (to fit the data) and $\ell_1$~norms (to
promote sparsity in wavelet space). 

In preparation of the continuing
increase in high-quality seismic data that is expected in the decades
to come, our focus has also been on numerical efficiency and the
ability to use parallel computing in constructing the model. We have
shown how our seismic model representation behaves under progressive
thresholding of the wavelet coefficients, and how the geographically
distributed power of published seismic models varies over the scale
lengths that can be independently resolved. Synthetic tests under
realistic conditions validates the approach that we advocate for the
future of seismic tomography, which shows the ability to explain
heterogeneous, massive data sets under the constraint that the
best-fitting models should also be sparse in the wavelet bases used.

\section*{A~C~K~N~O~W~L~E~D~G~E~M~E~N~T~S}

We thank Huub Douma and Massimo Fornasier for valuable discussions
throughout the past several years. I.~L. is ``chercheur qualifi\'e''
of the F.~R.~S.-FNRS.  (Belgium). Portions of this research were
supported by VUB-GOA grant 062 to I.~C.~D. and I.~L., the
FWO-Vlaanderen grant G.0564.09N to I.~C.~D.  and I.~L., and by NSF
grant CMG-0530865 to I.~C.~D., G.~N. and F.~A.~Dahlen. F.~J.~S. was
supported by Princeton University account 195-2142 while enjoying the
hospitality of the Vrije Universiteit Brussel and the Katholieke
Universiteit Leuven in the Summer of 2010, where the write-up started.
GN and JC received support from the ERC (Advanced grant 226837) and a
Marie Curie Re-integration Grant (project 223799). Computer code is
freely available from \url{homepages.ulb.ac.be/~igloris/} and
\url{www.frederik.net}.

\bibliography{bib}

\begin{thebibliography}{120}
\expandafter\ifx\csname natexlab\endcsname\relax\def\natexlab#1{#1}\fi

\bibitem[Aki et~al.(1977)Aki, Christoffersson, \& Husebye]{Aki+77}
Aki, K., Christoffersson, A. \& Husebye, E.~S., 1977.
Determination of the three-dimensional seismic structure of the lithosphere,
  {\it J.~Geophys.~Res.\/}, {\bf 82}(2), 277--296.

\bibitem[Amirbekyan \& Michel(2008)]{Amirbekyan+2008a}
Amirbekyan, A. \& Michel, V., 2008.
Splines on the 3-dimensional ball and their application to seismic body wave
  tomography, {\it Inverse Problems\/}, {\bf 24}, 015022, doi:
  10.1088/0266--5611/24/1/015022.

\bibitem[Amirbekyan et~al.(2008)Amirbekyan, Michel, \&
  Simons]{Amirbekyan+2008b}
Amirbekyan, A., Michel, V. \& Simons, F.~J., 2008.
Parameterizing surface-wave tomopgraphic models with harmonic spherical
  splines, {\it Geophys.~J.~Int.\/}, {\bf 174}(2), 617, doi:
  10.1111/j.1365--246X.2008.03809.x.

\bibitem[Antoine \& Vandergheynst(1999)]{Antoine+99}
Antoine, J.-P. \& Vandergheynst, P., 1999.
Wavelets on the 2-sphere: {A} group-theoretical approach, {\it
  Appl.~Comput.~Harmon.~Anal.\/}, {\bf 7}, 262--291.

\bibitem[Antoine et~al.(2002)Antoine, Demanet, Jacques, \&
  Vandergheynst]{Antoine+2002}
Antoine, J.-P., Demanet, L., Jacques, L. \& Vandergheynst, P., 2002.
Wavelets on the sphere: implementation and approximations, {\it
  Appl.~Comput.~Harmon.~Anal.\/}, {\bf 13}, 177--200.

\bibitem[Beck \& Teboulle(2009)]{Beck+2008}
Beck, A. \& Teboulle, M., 2009.
A fast iterative shrinkage-threshold algorithm for linear inverse problems,
  {\it SIAM J.~Imag.~Sci.\/}, {\bf 2}, 183--202, doi: 10.1137/080716542.

\bibitem[Becker \& Boschi(2002)]{Becker+2002}
Becker, T.~W. \& Boschi, L., 2002.
A comparison of tomographic and geodynamic mantle models, {\it
  Geochem.~Geophys.~Geosys.\/}, {\bf 3}, 1003.

\bibitem[Becker et~al.(2006)Becker, Chevrot, Schulte-Pelkum, \&
  Blackman]{Becker2006}
Becker, T.~W., Chevrot, S., Schulte-Pelkum, V. \& Blackman, D.~K., 2006.
Statistical properties of seismic anisotropy predicted by upper mantle
  geodynamic models, {\it J.~Geophys.~Res.\/}, {\bf 111}(B10), B08309,
  doi:10.1029/2005JB004095.

\bibitem[Becker et~al.(2007)Becker, Browaeys, \& Jordan]{Becker+2007}
Becker, T.~W., Browaeys, J.~T. \& Jordan, T.~H., 2007.
Stochastic analysis of shear-wave splitting length scales, {\it
  Earth~Planet.~Sci.~Lett.\/}, {\bf 259}(3--4), 526--540.

\bibitem[Bergeron et~al.(1999)Bergeron, Vincent, Yuen, Tranchant, \&
  Tchong]{Bergeron+99}
Bergeron, S.~Y., Vincent, A.~P., Yuen, D.~A., Tranchant, B. J.~S. \& Tchong,
  C., 1999.
Viewing seismic velocity anomalies with 3-{D} continuous {G}aussian wavelets,
  {\it Geophys.~Res.~Lett.\/}, {\bf 26}(15), 2311--2314.

\bibitem[Blumensath \& Davies(2008)]{Blumensath.Davies2008}
Blumensath, T. \& Davies, M.~E., 2008.
Iterative thresholding for sparse approximations, {\it J.~Fourier
  Anal.~Appl.\/}, {\bf 14}(5, doi: 10.1007/s00041-008-9035-z), 629--654.

\bibitem[Blumensath \& Davies(2009)]{Blumensath.Davies2009a}
Blumensath, T. \& Davies, M.~E., 2009.
Iterative hard thresholding for compressed sensing, {\it
  Appl.~Comput.~Harmon.~Anal.\/}, {\bf 27}(3), 265--274.

\bibitem[Boschi \& Dziewo{\'n}ski(1999)]{Boschi+99}
Boschi, L. \& Dziewo{\'n}ski, A.~M., 1999.
High- and low-resolution images of the {E}arth's mantle. {I}mplications of
  different approaches to tomographic modeling, {\it J.~Geophys.~Res.\/}, {\bf
  104}(B11), 25567--25594.

\bibitem[Boschi et~al.(2004)Boschi, Ekstr{\"o}m, \& Kustowksi]{Boschi+2004}
Boschi, L., Ekstr{\"o}m, G. \& Kustowksi, B., 2004.
Multiple resolution surface wave tomography: {t}he {M}editerranean basin, {\it
  Geophys.~J.~Int.\/}, {\bf 157}, 293--304, doi:
  10.1111/j.1365--246X.2004.02194.x.

\bibitem[Bozda{\u{g}} \& Trampert(2010)]{Bozdag+2010}
Bozda{\u{g}}, E. \& Trampert, J., 2010.
Assessment of tomographic mantle models using spectral element seismograms,
  {\it Geophys.~J.~Int.\/}, {\bf 180}(3), 1187--1199.

\bibitem[Bruckstein et~al.(2009)Bruckstein, Donoho, \& Elad]{Bruckstein+2009}
Bruckstein, A.~M., Donoho, D.~L. \& Elad, M., 2009.
From sparse solutions of systems of equations to sparse modeling of signals and
  images, {\it SIAM Rev.\/}, {\bf 51}(1), 34--81, doi: 10.1137/060657704.

\bibitem[Bull et~al.(2009)Bull, McNamara, \& Ritsema]{Bull+2009}
Bull, A.~L., McNamara, A.~K. \& Ritsema, J., 2009.
Synthetic tomography of plume clusters and thermochemical piles, {\it
  Earth~Planet.~Sci.~Lett.\/}, {\bf 278}(3--4), 152--162.

\bibitem[Cammarano et~al.(2005)Cammarano, Goes, Deuss, \&
  Giardini]{Cammarano+2005}
Cammarano, F., Goes, S., Deuss, A. \& Giardini, D., 2005.
Is a pyrolitic adiabatic mantle compatible with seismic data?, {\it
  Earth~Planet.~Sci.~Lett.\/}, {\bf 232}(3--4), 227--243, doi:
  10.1016/j.epsl.2005.01.03.

\bibitem[Cand{\`e}s et~al.(2005)Cand{\`e}s, Demanet, Donoho, \&
  Ying]{Candes+2005b}
Cand{\`e}s, E.~J., Demanet, L., Donoho, D.~L. \& Ying, L., 2005.
Fast discrete curvelet transforms, {\it Multisc.~Model.~Simul.\/}, {\bf 5},
  861--899.

\bibitem[Cand{\`e}s et~al.(2006)Cand{\`e}s, Romberg, \& Tao]{Candes+2006}
Cand{\`e}s, E.~J., Romberg, J.~K. \& Tao, T., 2006.
Stable signal recovery from incomplete and inaccurate measurements, {\it
  Comm.~Pure\ Appl.~Math.\/}, {\bf 59}(8), 1207--1223.

\bibitem[Chambolle \& Lions(1997)]{Chambolle+97}
Chambolle, A. \& Lions, P.-L., 1997.
Image recovery via total variation minimization and related problems, {\it
  Numer.~Math.\/}, {\bf 76}(2), 167--188.

\bibitem[Chevrot \& Zhao(2007)]{Chevrot+2007}
Chevrot, S. \& Zhao, L., 2007.
Multiscale finite-frequency {R}ayleigh wave tomography of the {K}aapvaal
  craton, {\it Geophys.~J.~Int.\/}, {\bf 169}(1), 201--215, doi:
  10.1111/j.1365--246X.2006.03289.x.

\bibitem[Chevrot et~al.(1998{\natexlab{a}})Chevrot, Montagner, \&
  Snieder]{Chevrot+98a}
Chevrot, S., Montagner, J.-P. \& Snieder, R.~K., 1998.
The spectrum of tomographic {E}arth models, {\it Geophys.~J.~Int.\/}, {\bf
  133}, 783--788.

\bibitem[Chevrot et~al.(1998{\natexlab{b}})Chevrot, Montagner, \&
  Snieder]{Chevrot+98b}
Chevrot, S., Montagner, J.-P. \& Snieder, R.~K., 1998.
The spectrum of tomographic {E}arth models: {C}orrection, {\it
  Geophys.~J.~Int.\/}, {\bf 135}, 311.

\bibitem[Chiao \& Liang(2003)]{Chiao+2003}
Chiao, L. \& Liang, W.~T., 2003.
Multiresolution parameterization for geophysical inverse problems, {\it
  Geophysics\/}, {\bf 68}, 199.

\bibitem[Chiao \& Kuo(2001)]{Chiao+2001}
Chiao, L.-Y. \& Kuo, B.-Y., 2001.
Multiscale seismic tomography, {\it Geophys.~J.~Int.\/}, {\bf 145}, 517--527,
  10.1046/j.0956--540x.2001.01403.x.

\bibitem[Cohen et~al.(1992)Cohen, Daubechies, \& Feauveau]{Cohen+92}
Cohen, A., Daubechies, I. \& Feauveau, J., 1992.
Biorthogonal bases of compactly supported wavelets, {\it Comm.~Pure\
  Appl.~Math.\/}, {\bf 45}, 485--560, doi: 10.1002/cpa.3160450502.

\bibitem[Cohen et~al.(1993)Cohen, Daubechies, \& Vial]{Cohen+93}
Cohen, A., Daubechies, I. \& Vial, P., 1993.
Wavelets on the interval and fast wavelet transforms, {\it
  Appl.~Comput.~Harmon.~Anal.\/}, {\bf 1}, 54--81.

\bibitem[Constable et~al.(1987)Constable, Parker, \& Constable]{Constable+87}
Constable, S.~C., Parker, R.~L. \& Constable, C.~G., 1987.
Occam's inversion: {A} practical algorithm for generating smooth models from
  electromagnetic sounding data, {\it Geophysics\/}, {\bf 52}(3), 289--300.

\bibitem[Daubechies(1988)]{Daubechies88b}
Daubechies, I., 1988.
Orthonormal bases of compactly supported wavelets, {\it Comm.~Pure\
  Appl.~Math.\/}, {\bf 41}, 909--996.

\bibitem[Daubechies(1992)]{Daubechies92}
Daubechies, I., 1992.
{\it Ten Lectures on Wavelets\/}, vol.~61 of {\bf CBMS-NSF Regional Conference
  Series in Applied Mathematics}, Society for Industrial \& Applied
  Mathematics, Philadelphia, Penn.

\bibitem[Daubechies et~al.(2004)Daubechies, Defrise, \& Mol]{Daubechies+2004}
Daubechies, I., Defrise, M. \& Mol, C.~D., 2004.
An iterative thresholding algorithm for linear inverse problems with a sparsity
  constraint, {\it Comm.~Pure\ Appl.~Math.\/}, {\bf 57}(11), 1413--1457, doi:
  10.1002/cpa.20042.

\bibitem[Debayle \& Sambridge(2004)]{Debayle+2004}
Debayle, E. \& Sambridge, M., 2004.
Inversion of massive surface wave data sets: {M}odel construction and
  resolution assessment, {\it J.~Geophys.~Res.\/}, {\bf 109}, B02316, doi:
  10.1029/2003JB002652.

\bibitem[Dobson \& Santosa(1996)]{Dobson+96}
Dobson, D.~C. \& Santosa, F., 1996.
Recovery of blocky images from noisy and blurred data, {\it SIAM
  J.~Appl.~Math.\/}, {\bf 56}(4), 1181--1198.

\bibitem[Donoho(2006)]{Donoho2006}
Donoho, D.~L., 2006.
For most large underdetermined systems of linear equations the minimal
  $\ell_1$-norm solution is also the sparsest solution, {\it Comm.~Pure\
  Appl.~Math.\/}, {\bf 59}(6), 797--829, doi: 10.1002/cpa.20132.

\bibitem[Donoho \& Johnstone(1994)]{Donoho+94}
Donoho, D.~L. \& Johnstone, I.~M., 1994.
Ideal spatial adaptation by wavelet shrinkage, {\it Biometrika\/}, {\bf 81}(3),
  425--455.

\bibitem[Donoho \& Johnstone(1995)]{Donoho+95b}
Donoho, D.~L. \& Johnstone, I.~M., 1995.
Adapting to unknown smoothness via wavelet shrinkage, {\it
  J.~Acoust.~Soc.~Am.\/}, {\bf 90}(432).

\bibitem[Dziewo{\'n}ski(1984)]{Dziewonski84}
Dziewo{\'n}ski, A.~M., 1984.
Mapping the lower mantle: {D}etermination of lateral heterogeneity in
  \textit{{P}} velocity up to degree and order~6, {\it J.~Geophys.~Res.\/},
  {\bf 89}(B7), 5929--5952.

\bibitem[Easley et~al.(2008)Easley, Lim, \& Labate]{Easley+2008}
Easley, G., Lim, W. \& Labate, D., 2008.
Sparse directional image representations using the discrete shearlet transform,
  {\it Appl.~Comput.~Harmon.~Anal.\/}, {\bf 25}, 25--46.

\bibitem[Efron et~al.(2004)Efron, Hastie, Johnstone, \& Tibshirani]{EfHJT2004}
Efron, B., Hastie, T., Johnstone, I. \& Tibshirani, R., 2004.
Least angle regression, {\it Ann.~Statist.\/}, {\bf 32}(2), 407--499, doi:
  10.1214/009053604000000067.

\bibitem[Ekstr{\"o}m et~al.(1997)Ekstr{\"o}m, Tromp, \& Larson]{Ekstrom+97}
Ekstr{\"o}m, G., Tromp, J. \& Larson, E. W.~F., 1997.
Measurements and global models of surface wave propagation, {\it
  J.~Geophys.~Res.\/}, {\bf 102}(B4), 8137--8157.

\bibitem[Foufoula-Georgiou \& Kumar(1994)]{Foufoula+94}
Foufoula-Georgiou, E. \& Kumar, P., eds., 1994.
{\it Wavelets in Geophysics\/}, Academic Press, San Diego, Calif.

\bibitem[Freeden \& Michel(1999)]{Freeden+99}
Freeden, W. \& Michel, V., 1999.
Constructive approximation and numerical methods in geodetic research today --
  an attempt at a categorization based on an uncertainty principle, {\it
  J.~Geodesy\/}, {\bf 73}(9), 452--465.

\bibitem[Freeden \& Michel(2004{\natexlab{a}})]{Freeden+2004a}
Freeden, W. \& Michel, V., 2004.
Orthogonal zonal, tesseral and sectorial wavelets on the sphere for the
  analysis of satellite data, {\it Adv.~Comput.~Math.\/}, {\bf 21}(1--2),
  181--217.

\bibitem[Freeden \& Michel(2004{\natexlab{b}})]{Freeden+2004b}
Freeden, W. \& Michel, V., 2004.
{\it Multiscale potential theory\/}, Birkh\"auser, Boston, Mass.

\bibitem[Garcia et~al.(2009)Garcia, Chevrot, \& Calvet]{Garcia+2009}
Garcia, R.~F., Chevrot, S. \& Calvet, M., 2009.
Statistical study of seismic heterogeneities at the base of the mantle from
  \textit{{PKP}} differential traveltimes, {\it Geophys.~J.~Int.\/}, {\bf
  179}(3), 1607--1616.

\bibitem[Gauch(2003)]{Gauch2003}
Gauch, H.~G., 2003.
{\it Scientific method in practice\/}, Cambridge Univ.~Press, Cambridge, UK.

\bibitem[Gholami \& Siahkoohi(2010)]{Gholami+2010}
Gholami, A. \& Siahkoohi, H.~R., 2010.
Regularization of linear and non-linear geophysical ill-posed problems with
  joint sparsity constraints, {\it Geophys.~J.~Int.\/}, {\bf 180}(2), 871--882,
  doi: 10.1111/j.1365--246X.2009.04453.x.

\bibitem[Gurnis(1986)]{Gurnis86}
Gurnis, M., 1986.
Quantitative bounds on the size spectrum of isotopic heterogeneity within the
  mantle, {\it Nature\/}, {\bf 323}, 317--320, doi:10.1038/323317a0.

\bibitem[Hansen(1992)]{Hansen92}
Hansen, P.~C., 1992.
Analysis of discrete ill-posed problems by means of the {L}-curve, {\it SIAM
  Rev.\/}, {\bf 34}(4), 561--580, doi: 10.1137/1034115.

\bibitem[Hedlin \& Shearer(2000)]{Hedlin+2000}
Hedlin, M. A.~H. \& Shearer, P.~M., 2000.
An analysis of large-scale variations in small-scale mantle heterogeneity using
  {G}lobal {S}eismographic {N}etwork recordings of precursors to
  \textit{{PKP}}, {\it J.~Geophys.~Res.\/}, {\bf 105}(B6), 13655.

\bibitem[Hemmat et~al.(2005)Hemmat, Dehghan, \& Skopina]{Hemmat+2005}
Hemmat, A.~A., Dehghan, M.~A. \& Skopina, M., 2005.
Ridge wavelets on the ball, {\it J.~Approx.~Theory\/}, {\bf 136}(2), 129--139.

\bibitem[Hernlund \& Houser(2008)]{Hernlund+2008}
Hernlund, J.~W. \& Houser, C., 2008.
On the statistical distribution of seismic velocities in {E}arth's deep mantle,
  {\it Earth~Planet.~Sci.~Lett.\/}, {\bf 265}(3--4), 423--437.

\bibitem[Holschneider et~al.(2003)Holschneider, Chambodut, \&
  Mandea]{Holschneider+2003}
Holschneider, M., Chambodut, A. \& Mandea, M., 2003.
From global to regional analysis of the magnetic field on the sphere using
  wavelet frames, {\it Phys.~Earth Planet.~Inter.\/}, {\bf 135}, 107--124.

\bibitem[Houser \& Williams(2009)]{Houser+2009}
Houser, C. \& Williams, Q., 2009.
The relative wavelengths of fast and slow velocity anomalies in the lower
  mantle: {C}ontrary to the expectations of dynamics?, {\it Phys.~Earth
  Planet.~Inter.\/}, {\bf 176}(3--04), 187--197.

\bibitem[Hung et~al.(2010)Hung, Chen, Chiao, \& Tseng]{Hung+2010}
Hung, S.-H., Chen, W.-P., Chiao, L.-Y. \& Tseng, T.-L., 2010.
First multi-scale, finite-frequency tomography illuminates {3D} anatomy of the
  {T}ibetan {P}lateau, {\it Geophys.~Res.~Lett.\/}, {\bf 37}, L06304,
  doi:10.1029/2009GL041875.

\bibitem[Jawerth \& Sweldens(1994)]{Jawerth+94}
Jawerth, B. \& Sweldens, W., 1994.
An overview of wavelet-based multiresolution analyses, {\it SIAM Rev.\/}, {\bf
  36}(3), 377--412.

\bibitem[Jensen \& la~Cour-Harbo(2001)]{Jensen+2001}
Jensen, A. \& la~Cour-Harbo, A., 2001.
{\it Ripples {i}n Mathematics\/}, Springer, Berlin.

\bibitem[Jordan et~al.(1993)Jordan, Puster, Glatzmaier, \& Tackley]{Jordan+93}
Jordan, T.~H., Puster, P., Glatzmaier, G.~A. \& Tackley, P.~J., 1993.
Comparisons between seismic {E}arth structures and mantle flow models based on
  radial correlation functions, {\it Science\/}, {\bf 261}(5127), 1427--1431.

\bibitem[K{\'a}rason \& van~der Hilst(2000)]{Karason+2000}
K{\'a}rason, H. \& van~der Hilst, R.~D., 2000, Constraints on mantle convection
  from seismic tomography, in {\em The History {a}nd Dynamics {o}f Global Plate
  Motions\/}, edited by M.~A. Richards, R.~G. Gordon, \& R.~D. van~der Hilst,
  vol. 121 of {\bf Geophysical Monograph}, Amer.~Geophys.~Union, Washington,
  D.~C.

\bibitem[Klees \& Haagmans(2000)]{Klees+2000}
Klees, R. \& Haagmans, R. H.~N., eds., 2000.
{\it Wavelets in the {G}eosciences\/}, vol.~90 of {\bf Lecture Notes in Earth
  Sciences}, Springer, Berlin.

\bibitem[Komatitsch \& Tromp(2002)]{Komatitsch+2002a}
Komatitsch, D. \& Tromp, J., 2002.
Spectral-element simulations of global seismic wave propagation --- {I}.
  {V}alidation, {\it Geophys.~J.~Int.\/}, {\bf 149}, 390--412.

\bibitem[Labate et~al.(2005)Labate, Lim, Kutyniok, \& Weiss]{Labate+2005}
Labate, D., Lim, W.-Q., Kutyniok, G. \& Weiss, G., 2005, Sparse
  multidimensional representation using shearlets, in {\em Wavelets~{XI}\/},
  edited by M.~Papadakis, A.~F. Laine, \& M.~A. Unser, vol. 5914, pp. 254--262,
  doi: 10.1117/12.613494, SPIE.

\bibitem[Lessig \& Fiume(2008)]{Lessig+2008}
Lessig, C. \& Fiume, E., 2008.
{SOHO}: Orthogonal and symmetric {H}aar wavelets on the sphere, {\it ACM
  Trans.~Graph.\/}, {\bf 27}(1), 4, doi: 10.1145/1330511.1330515.

\bibitem[Loris(2008)]{Loris2008}
Loris, I., 2008.
{L1Packv2}: A {M}athematica package for minimizing an $\ell_1$-penalized
  functional, {\it Comput.~Phys.~Comm.\/}, {\bf 179}, 895--902, doi:
  10.1016/j.cpc.2008.07.010.

\bibitem[Loris(2009)]{Loris2009}
Loris, I., 2009.
On the performance of algorithms for the minimization of $\ell_1$-penalized
  functionals, {\it Inv.~Probl.\/}, {\bf 25}, 035008,
  doi:10.1088/0266--5611/25/3/035008.

\bibitem[Loris et~al.(2007)Loris, Nolet, Daubechies, \& Dahlen]{Loris+2007}
Loris, I., Nolet, G., Daubechies, I. \& Dahlen, F.~A., 2007.
Tomographic inversion using $\ell_1$-norm regularization of wavelet
  coefficients, {\it Geophys.~J.~Int.\/}, {\bf 170}(1), 359--370, doi:
  10.1111/j.1365--246X.2007.03409.x.

\bibitem[Loris et~al.(2010)Loris, Douma, Nolet, Daubechies, \&
  Regone]{Loris+2010}
Loris, I., Douma, H., Nolet, G., Daubechies, I. \& Regone, C., 2010.
Nonlinear regularization techniques for seismic tomography, {\it
  J.~Comput.~Phys.\/}, {\bf 229}(3), 890--905, doi: 10.1016/j.jcp.2009.10.020.

\bibitem[Mallat(2008)]{Mallat2008}
Mallat, S., 2008.
{\it A Wavelet Tour {o}f Signal Processing, {T}he Sparse Way\/}, Academic
  Press, San Diego, Calif., 3rd edn.

\bibitem[Mallat(1989)]{Mallat89a}
Mallat, S.~G., 1989.
Multiresolution approximations and wavelet orthonormal bases of {L}2({R}), {\it
  Trans.~Am.~Math.~Soc.\/}, {\bf 315}(1), 69--87.

\bibitem[Margerin \& Nolet(2003)]{Margerin+2003}
Margerin, L. \& Nolet, G., 2003.
{Multiple scattering of high-frequency seismic waves in the deep {E}arth:
  \textit{{PKP}} precursor analysis and inversion for mantle granularity}, {\it
  J.~Geophys.~Res.\/}, {\bf 108}, 2514.

\bibitem[McEwen et~al.(2007)McEwen, Hobson, Mortlock, \& Lasenby]{McEwen+2007}
McEwen, J.~D., Hobson, M.~P., Mortlock, D.~J. \& Lasenby, A.~N., 2007.
Fast directional continuous spherical wavelet transform algorithms, {\it IEEE
  Trans.~Signal~Process.\/}, {\bf 55}(2), 520--529.

\bibitem[M{\'e}gnin et~al.(1997)M{\'e}gnin, Bunge, Romanowicz, \&
  Richards]{Megnin+97}
M{\'e}gnin, C., Bunge, H.-P., Romanowicz, B. \& Richards, M.~A., 1997.
Imaging {3-D} spherical convection models: {W}hat can seismic tomography tell
  us about mantle dynamics?, {\it Geophys.~Res.~Lett.\/}, {\bf 24}(11),
  1299--1302, doi: 10.1029/97GL01256.

\bibitem[Montelli et~al.(2004)Montelli, Nolet, Dahlen, Masters, Engdahl, \&
  Hung]{Montelli+2004b}
Montelli, R., Nolet, G., Dahlen, F.~A., Masters, G., Engdahl, E.~R. \& Hung,
  S.-H., 2004.
Global \textit{{P}} and \textit{{PP}} traveltime tomography: rays versus waves,
  {\it Geophys.~J.~Int.\/}, {\bf 158}, 637--654.

\bibitem[Montelli et~al.(2006)Montelli, Nolet, Dahlen, \&
  Masters]{Montelli+2006}
Montelli, R., Nolet, G., Dahlen, F.~A. \& Masters, G., 2006.
A catalogue of deep mantle plumes: new results from finite-frequency
  tomography, {\it Geochem.~Geophys.~Geosys.\/}, {\bf 7}, Q11007, doi:
  10.1029/2006GC001248.

\bibitem[Narcowich \& Ward(1996)]{Narcowich+96}
Narcowich, F.~J. \& Ward, J.~D., 1996.
Nonstationary wavelets on the m-sphere for scattered data, {\it
  Appl.~Comput.~Harmon.~Anal.\/}, {\bf 3}, 324--336.

\bibitem[Ni \& Helmberger(2003)]{Ni+2003}
Ni, S. \& Helmberger, D.~V., 2003.
Seismological constraints on the {S}outh {A}frican superplume; could be the
  oldest distinct structure on {E}arth, {\it EPSL\/}, {\bf 206}(1--2),
  119--131.

\bibitem[Nolet(1987)]{Nolet87}
Nolet, G., ed., 1987.
{\it Seismic Tomography\/}, Reidel, Hingham, MA.

\bibitem[Nolet(2008)]{Nolet2008}
Nolet, G., 2008.
{\it A Breviary for Seismic Tomography\/}, Cambridge Univ.~Press, Cambridge,
  UK.

\bibitem[Nolet \& Montelli(2005)]{Nolet+2005a}
Nolet, G. \& Montelli, R., 2005.
Optimal parametrization of tomographic models, {\it Geophys.~J.~Int.\/}, {\bf
  161}(2), 365--372, doi: 10.1111/j.1365--246X.2005.02596.x.

\bibitem[Oliver(2009)]{Oliver2009}
Oliver, M.~A., 2009.
Special issue {o}n applications {o}f wavelets {i}n the geosciences, {\it
  Math.~Geosc.\/}, {\bf 41}(6), 609--610.

\bibitem[Passier \& Snieder(1995)]{Passier+95}
Passier, M.~L. \& Snieder, R.~K., 1995.
On the presence of intermediate-scale heterogeneities in the upper mantle, {\it
  Geophys.~J.~Int.\/}, {\bf 123}, 817--837.

\bibitem[Piazzoni et~al.(2007)Piazzoni, Steinle-Neumann, Bunge, \&
  Dolej{\v{s}}]{Piazzoni+2007}
Piazzoni, A.~S., Steinle-Neumann, G., Bunge, H.-P. \& Dolej{\v{s}}, D., 2007.
A mineralogical model for density and elasticity of the {E}arth’s mantle,
  {\it Geochem.~Geophys.~Geosys.\/}, {\bf 8}, Q11010, doi:
  10.1029/2007GC001697.

\bibitem[Piromallo et~al.(2001)Piromallo, Vincent, Yuen, \&
  Morelli]{Piromallo+2001}
Piromallo, C., Vincent, A.~P., Yuen, D.~A. \& Morelli, A., 2001.
Dynamics of the transition zone under {E}urope inferred from wavelet
  cross-spectra of seismic tomography, {\it Phys.~Earth Planet.~Inter.\/}, {\bf
  125}, 125--139.

\bibitem[Press et~al.(1992)Press, Teukolsky, Vetterling, \& Flannery]{Press+92}
Press, W.~H., Teukolsky, S.~A., Vetterling, W.~T. \& Flannery, B.~P., 1992.
{\it Numerical Recipes {i}n {FORTRAN}: {T}he Art {o}f Scientific Computing\/},
  Cambridge Univ.~Press, New York, 2nd edn.

\bibitem[Puster et~al.(1995)Puster, Jordan, \& Hager]{Puster+95}
Puster, P., Jordan, T.~H. \& Hager, B.~H., 1995.
Characterization of mantle convection experiments using two-point correlation
  functions, {\it J.~Geophys.~Res.\/}, {\bf 100}(B4), 6351--6365.

\bibitem[Qin et~al.(2009)Qin, Capdeville, Montagner, Boschi, \&
  Becker]{Qin+2009}
Qin, Y., Capdeville, Y., Montagner, J.~P., Boschi, L. \& Becker, T.~W., 2009.
Reliability of mantle tomography models assessed by spectral element
  simulation, {\it Geophys.~J.~Int.\/}, {\bf 177}(1), 125--144.

\bibitem[Ritsema(2005)]{Ritsema2005}
Ritsema, J., 2005.
Global seismic structure maps, {\it Geol.~Soc.~Am.~Spec.~Paper\/}, {\bf 388},
  11--18, doi: 10.1130/2005.2388(02).

\bibitem[Ritsema et~al.(1999)Ritsema, van Heijst, \& Woodhouse]{Ritsema+99}
Ritsema, J., van Heijst, H.-J. \& Woodhouse, J.~H., 1999.
Complex shear wave velocity structure imaged beneath {A}frica and {I}celand,
  {\it Science\/}, {\bf 286}, 1925--1928.

\bibitem[Ritsema et~al.(2004)Ritsema, van Heijst, \& Woodhouse]{Ritsema+2004}
Ritsema, J., van Heijst, H.~J. \& Woodhouse, J.~H., 2004.
Global transition zone tomography, {\it J.~Geophys.~Res.\/}, {\bf 109}(B2),
  B02302, doi: 10.1029/2003JB002610.

\bibitem[Ritsema et~al.(2007)Ritsema, McNamara, \& Bull]{Ritsema+2007}
Ritsema, J., McNamara, A.~K. \& Bull, A.~L., 2007.
Tomographic filtering of geodynamic models: {I}mplications for model
  interpretation and large-scale mantle structure, {\it J.~Geophys.~Res.\/},
  {\bf 112}, B01303, doi: 10.1029/2006JB004566.

\bibitem[Ritsema et~al.(2010)Ritsema, A.~Deuss, van Heijst, \&
  Woodhouse]{Ritsema+2010}
Ritsema, J., A.~Deuss, A., van Heijst, H.~J. \& Woodhouse, J.~H., 2010.
{S40RTS}: {a} degree-40 shear-velocity model for the mantle from new {R}ayleigh
  wave dispersion, teleseismic traveltime and normal-mode splitting function
  measurements, {\it Geophys.~J.~Int.\/}, pp. doi:
  10.1111/j.1365--246X.2010.04884.x.

\bibitem[Ronchi et~al.(1996)Ronchi, Iacono, \& Paolucci]{Ronchi+96}
Ronchi, C., Iacono, R. \& Paolucci, P.~S., 1996.
The ``{C}ubed {S}phere'': {A} new method for the solution of partial
  differential equations in spherical geometry, {\it J.~Comput.~Phys.\/}, {\bf
  124}, 93--114, doi: 10.1006/jcph.1996.0047.

\bibitem[Rudin et~al.(1992)Rudin, Osher, \& Fatemi]{Rudin+92}
Rudin, L.~I., Osher, S. \& Fatemi, E., 1992.
Nonlinear total variation based noise removal algorithms, {\it Physica {D}\/},
  {\bf 60}(1--4), 259--268, doi: 10.1016/0167--2789(92)90242--F.

\bibitem[Sambridge et~al.(2006)Sambridge, Beghein, Simons, \&
  Snieder]{Sambridge+2006}
Sambridge, M., Beghein, C., Simons, F.~J. \& Snieder, R., 2006.
How do we understand and visualize uncertainty?, {\it The Leading Edge\/}, {\bf
  25}(5), 542--546.

\bibitem[Schmidt et~al.(2006)Schmidt, Han, Kusche, Sanchez, \&
  Shum]{Schmidt+2006}
Schmidt, M., Han, S.-C., Kusche, J., Sanchez, L. \& Shum, C.~K., 2006.
Regional high-resolution spatiotemporal gravity modeling from {GRACE} data
  using spherical wavelets, {\it Geophys.~Res.~Lett.\/}, {\bf 33}(8), L0840,
  doi: 10.1029/2005GL025509.

\bibitem[Schr{\"o}der \& Sweldens(1995)]{Schroder+95}
Schr{\"o}der, P. \& Sweldens, W., 1995.
Spherical wavelets: {E}fficiently representing functions on the sphere, {\it
  Computer Graphics Proceedings (SIGGRAPH 95)\/}, pp. 161--172.

\bibitem[Schuberth et~al.(2009)Schuberth, Bunge, \& Ritsema]{Schuberth+2009}
Schuberth, B. S.~A., Bunge, H.-P. \& Ritsema, J., 2009.
Tomographic filtering of high-resolution mantle circulation models: Can seismic
  heterogeneity be explained by temperature alone?, {\it
  Geochem.~Geophys.~Geosys.\/}, {\bf 10}.

\bibitem[Shearer \& Earle(2004)]{Shearer+2004}
Shearer, P.~M. \& Earle, P.~S., 2004.
The global short-period wavefield modelled with a {M}onte {C}arlo seismic
  phonon method, {\it Geophys.~J.~Int.\/}, {\bf 158}(3), 1103--1117.

\bibitem[Simons et~al.(2002)Simons, van~der Hilst, Montagner, \&
  Zielhuis]{Simons+2002b}
Simons, F.~J., van~der Hilst, R.~D., Montagner, J.-P. \& Zielhuis, A., 2002.
Multimode {R}ayleigh wave inversion for heterogeneity and azimuthal anisotropy
  of the {A}ustralian upper mantle, {\it Geophys.~J.~Int.\/}, {\bf 151}(3),
  738--754, doi: 10.1046/j.1365--246X.2002.01787.x.

\bibitem[Simons et~al.(2009)Simons, Nolet, Georgief, Babcock, Regier, \&
  Davis]{Simons+2009a}
Simons, F.~J., Nolet, G., Georgief, P., Babcock, J.~M., Regier, L.~A. \& Davis,
  R.~E., 2009.
On the potential of recording earthquakes for global seismic tomography by
  low-cost autonomous instruments in the oceans, {\it J.~Geophys.~Res.\/}, {\bf
  114}, B05307, doi:10.1029/2008JB006088.

\bibitem[Spakman \& Bijwaard(2001)]{Spakman+2001}
Spakman, W. \& Bijwaard, H., 2001.
Optimization of cell parameterization for tomographic inverse problems, {\it
  Pure Appl.~Geophys.\/}, {\bf 158}, 1401--1423.

\bibitem[Starck et~al.(2006)Starck, Moudden, Abrial, \& Nguyen]{Starck+2006}
Starck, J.~L., Moudden, Y., Abrial, P. \& Nguyen, M., 2006.
Wavelets, ridgelets and curvelets on the sphere, {\it Astron.~Astroph.\/}, {\bf
  446}, 1191--1204.

\bibitem[Strang \& Nguyen(1997)]{Strang+97}
Strang, G. \& Nguyen, T., 1997.
{\it Wavelets {a}nd Filter Banks\/}, Wellesley-Cambridge Press, Wellesley,
  Mass., 2nd edn.

\bibitem[Trampert \& Snieder(1996)]{Trampert+96b}
Trampert, J. \& Snieder, R., 1996.
Model estimations biased by truncated expansions: {P}ossible artifacts in
  seismic tomography, {\it Science\/}, {\bf 271}(5253), 1257--1260,
  doi:10.1126/science.271.5253.1257.

\bibitem[Trampert \& Woodhouse(1995)]{Trampert+95}
Trampert, J. \& Woodhouse, J.~H., 1995.
Global phase-velocity maps of {L}ove and {R}ayleigh-waves between 40 and 150
  seconds, {\it Geophys.~J.~Int.\/}, {\bf 122}(2), 675--690.

\bibitem[Trampert \& Woodhouse(1996)]{Trampert+96a}
Trampert, J. \& Woodhouse, J.~H., 1996.
High resolution global phase velocity distributions, {\it
  Geophys.~Res.~Lett.\/}, {\bf 23}(1), 21--24.

\bibitem[Trampert \& Woodhouse(2001)]{Trampert+2001}
Trampert, J. \& Woodhouse, J.~H., 2001.
Assessment of global phase velocity models, {\it Geophys.~J.~Int.\/}, {\bf
  144}(1), 165--174, doi: 10.1046/j.1365--246x.2001.00307.x.

\bibitem[van~der Hilst \& K{\'a}rason(1999)]{Hilst+99a}
van~der Hilst, R.~D. \& K{\'a}rason, H., 1999.
Compositional heterogeneity in the bottom 1000 kilometers of {E}arth's mantle:
  {T}oward a hybrid convection model, {\it Science\/}, {\bf 283}(5409),
  1885--1888.

\bibitem[Van{D}ecar \& Snieder(1994)]{VandeCar+94}
Van{D}ecar, J.~C. \& Snieder, R., 1994.
Obtaining smooth solutions to large, linear inverse problems, {\it
  Geophysics\/}, {\bf 59}(5), 818--829.

\bibitem[Wang \& Dahlen(1995)]{Wang+95a}
Wang, Z. \& Dahlen, F.~A., 1995.
Spherical-spline parameterization of three-dimensional {E}arth models, {\it
  Geophys.~Res.~Lett.\/}, {\bf 22}, 3099--3102.

\bibitem[Wang et~al.(1998)Wang, Tromp, \& Ekstr\"om]{Wang+98}
Wang, Z., Tromp, J. \& Ekstr\"om, G., 1998.
Global and regional surface-wave inversions: {A} spherical-spline
  parameterization, {\it Geophys.~Res.~Lett.\/}, {\bf 25}(2), 207--210.

\bibitem[Wiaux et~al.(2005)Wiaux, Jacques, \& Vandergheynst]{Wiaux+2005}
Wiaux, Y., Jacques, L. \& Vandergheynst, P., 2005.
Correspondence principle between spherical and {E}uclidean wavelets, {\it
  Astroph.~J.\/}, {\bf 632}, 15--28, doi: 10.1086/432926.

\bibitem[Wiaux et~al.(2007)Wiaux, McEwen, \& Vielva]{Wiaux+2007}
Wiaux, Y., McEwen, J.~D. \& Vielva, P., 2007.
Complex data processing: {F}ast wavelet analysis on the sphere, {\it J.~Fourier
  Anal.~Appl.\/}, {\bf 13}(4), 477--493, 10.1007/s00041--006--6917--9.

\bibitem[Woodhouse \& Dziewo{\'n}ski(1984)]{Woodhouse+84}
Woodhouse, J.~H. \& Dziewo{\'n}ski, A.~M., 1984.
Mapping the upper mantle: {T}hree-dimensional modeling of {E}arth structure by
  inversion of seismic waveforms, {\it J.~Geophys.~Res.\/}, {\bf 89}(B7),
  5953--5986.

\bibitem[Wysession(1996)]{Wysession96}
Wysession, M.~E., 1996.
Large-scale structure at the core-mantle boundary from diffracted waves, {\it
  Nature\/}, {\bf 382}, 244--248.

\bibitem[Wysession et~al.(1999)Wysession, Langenhorst, Fouch, Fischer,
  Al-{E}qabi, Shore, \& Clarke]{Wysession+99}
Wysession, M.~E., Langenhorst, A., Fouch, M.~J., Fischer, K.~M., Al-{E}qabi,
  G.~I., Shore, P.~J. \& Clarke, T.~J., 1999.
Lateral variations in compressional/shear velocities at the base of the mantle,
  {\it Science\/}, {\bf 284}, 120--125.

\bibitem[Yanovskaya \& Ditmar(1990)]{Yanovskaya+90}
Yanovskaya, T.~B. \& Ditmar, P.~G., 1990.
Smoothness criteria in surface wave tomography, {\it Geophys.~J.~Int.\/}, {\bf
  102}, 63--72.

\bibitem[Yuen et~al.(2002)Yuen, Vincent, Kido, \& Vecsey]{Yuen+2002}
Yuen, D.~A., Vincent, A.~P., Kido, M. \& Vecsey, L., 2002.
Geophysical applications of multidimensional filtering with wavelets, {\it Pure
  Appl.~Geophys.\/}, {\bf 159}(10), 2285--2309.

\bibitem[Zhang \& Tanimoto(1993)]{Zhang+93}
Zhang, Y.-S. \& Tanimoto, T., 1993.
High-resolution global upper-mantle structure and plate-tectonics, {\it
  J.~Geophys.~Res.\/}, {\bf 98}(B6), 9793--9823.

\end{thebibliography}
\bibliographystyle{gji}
\label{lastpage}

\end{document}